\documentclass{aa}
\usepackage{url}
\usepackage[breaklinks=true]{hyperref}
\usepackage{twoopt}
\usepackage[english]{babel}          % English language/hyphenation
\usepackage{pdflscape}

\usepackage{natbib}
\bibpunct{(}{)}{;}{a}{}{,} %% natbib format for A&A and ApJ

\usepackage[utf8]{inputenc}
\usepackage[normalem]{ulem}
\usepackage{siunitx}
\usepackage{amsmath, amssymb}
\usepackage[farskip=0pt]{subfig}

\usepackage{multirow,bigstrut,ctable}
\usepackage{rotating}
\usepackage[varg]{txfonts} % A&A recommended fonts
\usepackage{threeparttable}

\hypersetup{
    colorlinks,
    linkcolor={blue!50!black},
    citecolor={blue!50!black},
}

\newcommand{\kms}{km s$^{-1}$}

\newcommand{\wsc}{\texttt{wsclean}}
\newcommand{\dppp}{\texttt{DPPP}}

\newcommand{\casa}{\texttt{CASA}}
\newcommand{\degree}{$^{\circ}$}
\newcommand\Jyb{Jy~beam$^{-1}$}
\newcommand\Jyp{Jy~pixel$^{-1}$}
\newcommand\mjyb{mJy~beam$^{-1}$}
\newcommand{\asec}{$^{\prime\prime}$}
\newcommand{\amin}{$^{\prime}$}
\newcommand\Msun{M$_{\odot}$}

\newcommand\Hii{H~\textsc{ii}}
\newcommand\Nii{[N~{\sc ii}]}
\newcommand\disperse{{\tt DisPerSE}}
\newcommand\filchap{{\tt FilChaP}}
\newcommand\emuni{pc~cm$^{-6}$}
\newcommand\puni{K~cm$^{-3}$}
\newcommand\cmc{cm$^{-3}$}

\begin{document}

\title{Filamentary structures of ionized gas in Cygnus X}

\author{K.~L.~Emig\inst{1,2,}\thanks{Jansky Fellow of the National Radio Astronomy Observatory}, 
G.~J.~White\inst{3,4}, 
P.~Salas\inst{5}, 
R.~L.~Karim\inst{6}, 
R.~J.~van~Weeren\inst{2}, \\
P.~J.~Teuben\inst{6}, 
A.~Zavagno\inst{7,8},
P.~Chiu\inst{4},
M.~Haverkorn\inst{9},
J.~B.~R.~Oonk\inst{2,10,11},
E.~Orr\'u\inst{10}, \\
I.~M.~Polderman\inst{9}, 
W.~Reich\inst{12}, 
H.~J.~A.~R\"ottgering\inst{2}, and 
A.~G.~G.~M.~Tielens\inst{2}}

\authorrunning{Emig et al.}

\institute{National Radio Astronomy Observatory, 520 Edgemont Road, Charlottesville, VA 22903, USA \\ %1
\email{kemig@nrao.edu}
\and Leiden Observatory, Leiden University, P.O.~Box 9513, 2300 RA, Leiden, the Netherlands %2
\and Department of Physics and Astronomy, The Open University, Walton Hall, Milton Keynes, MK7 6AA, UK %3
\and RAL Space, STFC Rutherford Appleton Laboratory, Chilton, Didcot, Oxfordshire, OX11 0QX, UK %4
\and Green Bank Observatory, P.O. Box 2, Green Bank, WV 24944, USA %5
\and Department of Astronomy, University of Maryland, College Park, MD 20742, USA %6
\and Aix Marseille Univ, CNRS, CNES, LAM, Marseille, France %7
\and Institut Universitaire de France (IUF), Paris, France %8
\and Department of Astrophysics/IMAPP, Radboud University, PO Box 9010, 6500 GL, the Netherlands %9
\and Netherlands Institute for Radio Astronomy (ASTRON), Postbus 2, 7990 AA, Dwingeloo, the Netherlands %10
\and SURF/SURFsara, Science Park 140, 1098 XG Amsterdam, the Netherlands %11
\and Max-Planck-Institut f\"ur Radioastronomie, Auf dem H\"ugel 69, 53121 Bonn, Germany %12
}

% These dates will be filled out by the publisher
\date{Received ... / Accepted ...}

\abstract
% Context
{Ionized gas probes the influence of massive stars on their environment. The Cygnus X region ($d \sim 1.5$~kpc) is one of the most massive star forming complexes in our Galaxy, in which the Cyg OB2 association (age of 3-5 Myr and stellar mass $2 \times 10^{4}$~\Msun) has a dominant influence. }
%Aims
{We observe the Cygnus X region at 148~MHz using the Low Frequency Array (LOFAR) and take into account short-spacing information during image deconvolution. Together with data from the Canadian Galactic Plane Survey, we investigate the morphology, distribution, and physical conditions of low-density ionized gas in a 4\degree~$\times$~4\degree\ ($\sim$100~pc~$\times$~100~pc) region at a resolution of 2\amin\ (0.9~pc).}
% Methods
{The Galactic radio emission in the region analyzed is almost entirely thermal (free-free) at 148~MHz, with emission measures ($EM$) of $10^3 < EM~{\rm[pc~cm^{-6}]} < 10^6$. As filamentary structure is a prominent feature of the emission, we use \disperse\ and \filchap\ to identify filamentary ridges and characterize their radial ($EM$) profiles. }
% Results
{The distribution of radial profiles has a characteristic width of 4.3~pc and a power-law distribution ($\beta = -1.8 \pm 0.1$) in peak $EM$ down to our completeness limit of 4200~\emuni. The electron densities of the filamentary structure range between $10 \lesssim n_e~{\rm[cm^{-3}]} \lesssim 400$ with a median value of 35~cm$^{-3}$, remarkably similar to \Nii\ surveys of ionized gas.}
% Conclusions
{Cyg~OB2 may ionize at most two-thirds of the total ionized gas and the ionized gas in filaments. More than half of the filamentary structures are likely photoevaporating surfaces flowing into a surrounding diffuse ($\sim$5~\cmc) medium. However, this is likely not the case for all ionized gas ridges. A characteristic width in the distribution of ionized gas points to the stellar winds of Cyg~OB2 creating a fraction of the ionized filaments through swept-up ionized gas or dissipated turbulence.}


\maketitle

%%%%%%%%%%%%%%%%%%%%%%%%%%%%%%%%%%%%%%%%%%%%%%%%%%

%%%%%%%%%%%%%%%%% BODY OF PAPER %%%%%%%%%%%%%%%%%%

%%%%%%%%%%%%%%%%%%%%%%
\section{Introduction}
\label{sec:intro}

The interaction of massive stars with their environment has a profound impact on the evolution of galaxies \citep{Hopkins2014, Hopkins2018} through the collective effects of protostellar outflows \citep{Bally2016}, ionizing radiation \citep{Matzner2002}, stellar winds and supernovae \citep[SN; ][]{Yorke1989}. One way to investigate their impact is through photoionized gas, in which stars of mass $M_{\star} \gtrsim 7$~\Msun, all O types and earlier than B3, produce extreme ultraviolet (EUV) radiation of $E \geq 13.6$~eV capable of photoionizing hydrogen in the surrounding medium. Ionized gas is an important component of the interstellar medium (ISM) that reflects feedback processes associated with (i) ionizing radiation -- through photoionized gas, radiation pressure, and photoevaporation, but ionized gas can also reflect feedback by (ii) stellar winds and supernovae processes  -- through shocks, turbulence, and gas phase changes, (iii) turbulent dissipation, and (iv) gas accretion onto galaxies.  

Early in their lifetimes, massive stars dissociate and ionize their immediate environment. Dense \Hii\ regions form as pockets of ionized gas, increasing the thermal gas pressure within the molecular cloud by three orders of magnitude. The subsequent expansion of an \Hii\ region can mechanically unbind the parent cloud and induce turbulent motions \citep[e.g.,][]{Walch2012} in the interstellar medium (ISM). As an \Hii\ region expands, its volume density diminishes. Stellar winds and radiation pressure are also important contributors to the expansion of the ionized gas volume \citep[e.g.,][]{Pabst2019,Pabst2020,Olivier2021}. As a result of peculiar motion and/or inhomogeneities in the medium, within a few Myr \citep[e.g.,][]{Mezger1978}, the star, its photons and the gas it ionizes enter a surrounding low-density ($n_e \sim 0.1 - 100$~\cmc) medium. 

Within the plane of the Galaxy, photoionized gas is found in a variety of environments (and referred to with a variety of different names). Dense ($n_e > 10^3$~\cmc) ionized gas pervades \Hii\ regions of a few pc in size. 
When stellar winds are influential in an \Hii\ region, the medium stratifies with ionized gas at larger radii and hot gas filling the inner regions \citep{Weaver1977,Churchwell2006}. From leaky \Hii\ regions, ionizing photons escaping through porous material create (partially) ionized gas ($1 - 100$~\cmc) in the envelopes of \Hii\ regions. This diffuse ionized gas ($1-100$~\cmc) permeates to larger volumes in blister \Hii\ regions. Assisted by supernova explosions, massive stars create large excavated regions or plasma tunnels containing fully ionized gas ($1 - 10$~\cmc). Over-pressured --- thus denser, brighter, and more readily detected --- photoevaporating ionized gas (of smaller path length) is frequently observed in a number of these scenarios as ionization fronts propagate into local neutral material. Ionizing photons which escape in these scenarios provide a surplus of the ionizing photon budget required to maintain \citep[e.g.,][]{Reynolds1984} the pervasive \citep[volume filling factor $\phi \sim 0.25$,][]{Kulkarni1988} warm ionized component of the ISM (WIM), that resides in the Galactic plane ($n_e \sim 0.1$~\cmc) and which is characterized by large ($z \sim 1$~kpc) scale heights \citep[for a review see][]{Haffner2009}.

Blind surveys and large targeted samples with thermal radio continua \citep{Mezger1978, Murray2010}, radio recombination line emission \citep{Shaver1976a, Lockman1976, Anantharamaiah1985b,Anantharamaiah1986, Roshi2000, Heiles1996b, Alves2015}, FIR fine structure line emission from \Nii\ \citep{Bennett1994,Goldsmith2015} and pulsar dispersion measures \citep[][and references therein]{Berkhuijsen2006} have brought to light properties of low-density ionized gas within the Galactic plane. \cite{Mezger1978} estimated that 84\% of ionizing photons are emitted by O stars outside of compact \Hii\ regions, in gas characterized by densities of $n_e \approx 5 - 10$~\cmc\ dubbed extended low-density (ELD) \Hii\ gas. However, a number of different conclusions have been reached regarding the dominant origin of this gas: (i) envelopes of \Hii\ regions \citep{Shaver1976a, Anantharamaiah1986, McKee1997}, (ii) a pervasive component \citep{Heiles1996b} which may be a continuation of the WIM \citep{Bennett1994,Berkhuijsen2006}, and (iii) from just a handful of the most luminous star-forming regions in the Galaxy \citep{Murray2010}. Thus the volume filling factor of gas with these densities is unclear. High-resolution pinhole surveys of FIR fine structure lines \citep{Goldsmith2015, Pineda2019, Langer2021} trace somewhat denser gas (with mean values of $n_e \approx 30 - 40$~\cmc) and could plausibly be tied with these three possibilities. 

The Cygnus X region is a massive star-forming complex which displays filament-like structure in low-density ionized gas \citep{Wendker1991}. ``Cygnus X'' refers to a $\sim$10\degree\ wide region in the galactic plane with enhanced radio emission \citep{Piddington1952} -- for an overview of the region see \citet{Reipurth2008}. This coherent region of massive star formation lies at $\sim$1.5~kpc \citep{Schneider2006, Rygl2012} and dominates the observed emission, despite the view looking down a spiral arm in this direction. Open and massive OB associations are seen in this direction at a similar distance -- Cyg OB1, OB2, OB6, OB9 \citep{Uyaniker2001}. Yet (a subset of) the region still retains a large reservoir ($M > 10^6$~\Msun) of molecular gas \citep{Schneider2006, Schneider2011}. Massive post main-sequence stars indicate that star formation began in the region $\sim$15 Myr ago \citep{Comeron2012, Comeron2016, Comeron2020}. Though the nature of the rarefied region is debated, it is consistent with the superbubble formalism \citep{McKee1977} indicating that stellar winds and/or supernovae have contributed to excavating the parent cloud(s) \citep{Bochkarev1985, Ackermann2011a}.

Part of the region that is referred to as Cygnus X North \citep{Schneider2006} harbors Cyg~OB2 and other smaller clusters and associations, some of which are actively forming stars \citep{Cong1977, Odenwald1990, Comeron1999, Comeron2001, LeDuigou2002, Marston2004, Motte2007, Beerer2010, Panwar2020}. The Cyg OB2 association strongly influences the medium, largely characterized through pillars and globules \citep{Wright2012, Schneider2016b, Deb2018}. Cyg~OB2 has a total stellar mass of $M_{\star} = 1.7^{+0.4}_{-0.3} \times 10^{4}$~\Msun\ \citep{Wright2015}. It is not bound gravitationally and likely formed that way, in a relatively low density environment given its mass \citep{Wright2014}, during bursts of star formation 3 and 5 Myr ago \citep{Wright2010, Berlanas2020}.

In this article, we investigate low-density ionized gas in Cygnus X (North) with 148~MHz continuum observations using the Low Frequency Array \citep[LOFAR;][]{vanHaarlem2013}. We observe this region, encompassing 16 square degrees ($\sim$10~kpc$^2$) within a single pointing, near to the Cyg~OB2 association, as it contains a wide variety of evolutionary stages of star-formation with low-density ionized gas. We aim to investigate the (filamentary structure of) ionized gas and characterize the influence of massive stars on their environment through ionization feedback.

We adopt a distance of $d = 1.5 \pm 0.1$~kpc to the Cygnus X region and Cyg~OB2 following \citet[][ see their Sec. 2.2 for a detailed discussion]{Comeron2020}. After an analysis of molecular line emission linked the region as a coherent structure influenced by Cyg~OB2 \citep{Schneider2006,Schneider2016b}, the distance to the Cygnus X complex was determined as $1.40 \pm 0.08$~kpc through maser observations of massive stars \citep{Rygl2012}. A recent {\it Gaia} analysis \citep{Berlanas2019} places a main subgroup (80\% of the OB population) of Cyg~OB2 at $1.76^{+0.37}_{-0.26}$~kpc with a smaller subgroup at $1.35 \pm 0.2$~kpc. While a clear separation in distance distinguishes these two subgroups, \cite{Comeron2020} points out that a systematic offset in parallax may affect (both of) the distance determinations. At $d = 1.5$~kpc, the physical scale is 1\arcmin~$\approx 0.44$~pc.

%%%%%%%%%%%%%%%%%%%%%%%%%%%%%%%%%%%%%%%%
\section{Data}
\label{sec:data}

%%%%%%%%%%%%%%%%%%%%%%%%%%%%%%%%%%%%%%%%
\subsection{LOFAR observations \& data processing}
\label{ssec:lofarobs}

In this section we describe the data processing of
LOFAR interferometric observations of one pointing centered approximately on the massive star forming region DR~21 at $(\alpha_{\mathrm{RA}},\delta_{\mathrm{dec}}) = (309.5500^{\circ}, +42.0708^{\circ})$ in J2000 coordinates. These data were obtained with the high band antennas (HBA) covering 110--190 MHz on July 19, 2013 under project LC0\_032 (PI: G. White). Continuous frequency coverage was obtained between 126 -- 165 MHz and which we make use of in this analysis. 23 stations of the full Dutch array recorded data during these observations, which have a maximum baseline of 120~km (4\asec) and minimum baseline lengths of 70~m corresponding to largest angular scales of 96\arcmin\ (1.6\degree). The on-source integration time is 6~hours and 50~minutes. A plot of the $uv$ coverage of this observation can be found in the Appendix as Figure~\ref{fig:uvcov}.

We recorded data throughout the observation also in the direction of Cygnus A (299.8682\degree, 40.7339\degree), a bright ($\sim$10$^4$~Jy), well-modeled source 7.4\degree\ away from the target center. The flexibility of digitally pointing the LOFAR beam allows HBA data to be recorded towards multiple phase centers within the station beam of $\sim$20\degree. Because of the close proximity and simultaneous observation, we use Cygnus A as the primary calibrator source to derive flux, bandpass and phase calibrations to transfer to the target pointing.

%%%%%%%%%%%%%%%%%%%%%%%%%%%%%%%%%%%%%%%%
\subsubsection{Calibrating the visibilities}
\label{sssec:calibration}

We describe the calibration of the visibility data in this section. We derive flux, bandpass and phase calibrations of the primary calibrator pointing. We then apply these to the target pointing. Lastly, we mitigate side lobe contribution from Cygnus A to the target pointing by subtracting Cygnus A visibilities from the target visibilities; the trade-off to LOFAR's large field of view is its susceptibility to strong side lobe contamination.  

To begin processing of both the primary calibrator and target Measurement Sets (MS), we flag the correlations between the two ``ears'' of each HBA core station \citep{vanHaarlem2013} and flag for radio frequency interference (RFI) with \texttt{AOFlagger} \citep{Offringa2012} at full time and frequency resolution (0.763~kHz channel) using the default HBA strategy \citep{Offringa2012}. We also flag station CS013 as its antennas were not properly phased up at the time of the observations. Then we average (by a factor of 64) the data to a channel resolution of 48.8~kHz. \footnote{We run these steps using version 2.20\_2 of the Default Pipeline Processing Platform \citep[\dppp;][]{vanDiepen2018}. This version of \dppp\ outputs the data with minor updates to the metadata of the Measurement Set -- as these data were recorded in Cycle 0 of LOFAR's operation -- that are necessary in order to use later versions of \dppp. In subsequent processing, we make use of the most recent \dppp\ versions.}

For the primary calibrator, we solve for the diagonal (XX and YY) gain, phase and amplitude, at full frequency resolution with \dppp. We use a LOFAR model of Cygnus A at 150 MHz consisting of 33~000 components \citep[courtesy J. McKean;][]{McKean2016}. The solutions have cleanly converged, and they also indicate the ionosphere was relatively mild.  We apply only the phase solutions. Then, we solve again for the diagonal gain, phase and amplitude. Using the LOFAR Solutions Tool \citep[LoSoTo;][]{DeGasperin2019}, we flag spurious jumps in the amplitude solutions of each channel -- likely due to instrumental instabilities -- as well as any residual RFI, which are identified as 7$\sigma$ outliers averaged over a sliding window of 40 solution intervals in time. 

For the target pointing, we use \dppp\ to apply the phase solutions of the first iteration of gain calibration of the primary calibrator and both the amplitude and phase solutions of the second iteration. The time steps of the visibility data which have flagged amplitude solutions will be flagged as the solutions are applied. Phase offsets due to asynchronized --- on the order of $\mathcal{O}$(10 nanoseconds) --- station clocks of the remote stations \citep{VanWeeren2016} have been absorbed into the primary calibrator solutions. Furthermore, a calm ionosphere may be characterized by large patches of coherence in the total electron content \citep{Intema2009}, thus inducing similar phase errors (and Faraday rotation) towards the target and nearby primary calibrator. Phase (and amplitude) errors due to the ionosphere are absorbed into these solutions as well. 

In the next step, we use \dppp\ to subtract Cygnus A from the target visibilities because of the close proximity of Cygnus A to our pointing center. We phase shift the target visibilities to the location of Cygnus A and then predict the 33~000 source model of Cygnus A into the MODEL column. Next we smooth our visibilities with a baseline dependent averager \citep{DeGasperin2019} and solve for the diagonal (XX and YY) gain at a 195.3~kHz frequency resolution and 1 minute time resolution, attempting to smear the visibility structure of the target. We corrupt the MODEL column with the solutions. Then we subtract the corrupted MODEL from the DATA. We phase-shift these visibilities back to the target phase center of the observation. In this manner, we have subtracted the off-axis signal of Cygnus A from the data. We find this approach produces more coherent solutions with a larger number of stations converging than with the standard ``demixing'' \citep{vanDerTol2007} capabilities of \dppp.

At this point, the target data are well calibrated against a source which has a peak flux 1000 times brighter than the apparent flux of any source in the target field. With our science aim to investigate extended diffuse emission that is well characterized with low-resolution (2\arcmin) imaging, self-calibration tests do not significantly improve our short baseline calibration.

%%%%%%%%%%%%%%%%%%%%%%%%%%%%%%%%%%%%%%%%
\subsubsection{Constructing a short-spacing map }
\label{sssec:ss}

The LOFAR data have excellent $uv$ coverage to large scale emission with baselines as short as 70~m providing sensitivity to a largest angular size of 96\arcmin. However, for observations near the Galactic plane, where large scale emission surpasses tens of degrees, information on the smallest $uv$ scales is necessary to properly deconvolve and obtain the flux density of the extended $\mathcal{O}$(1\degree) structures in the Cygnus X region.

\begin{figure}
    \includegraphics[width=0.48\textwidth]{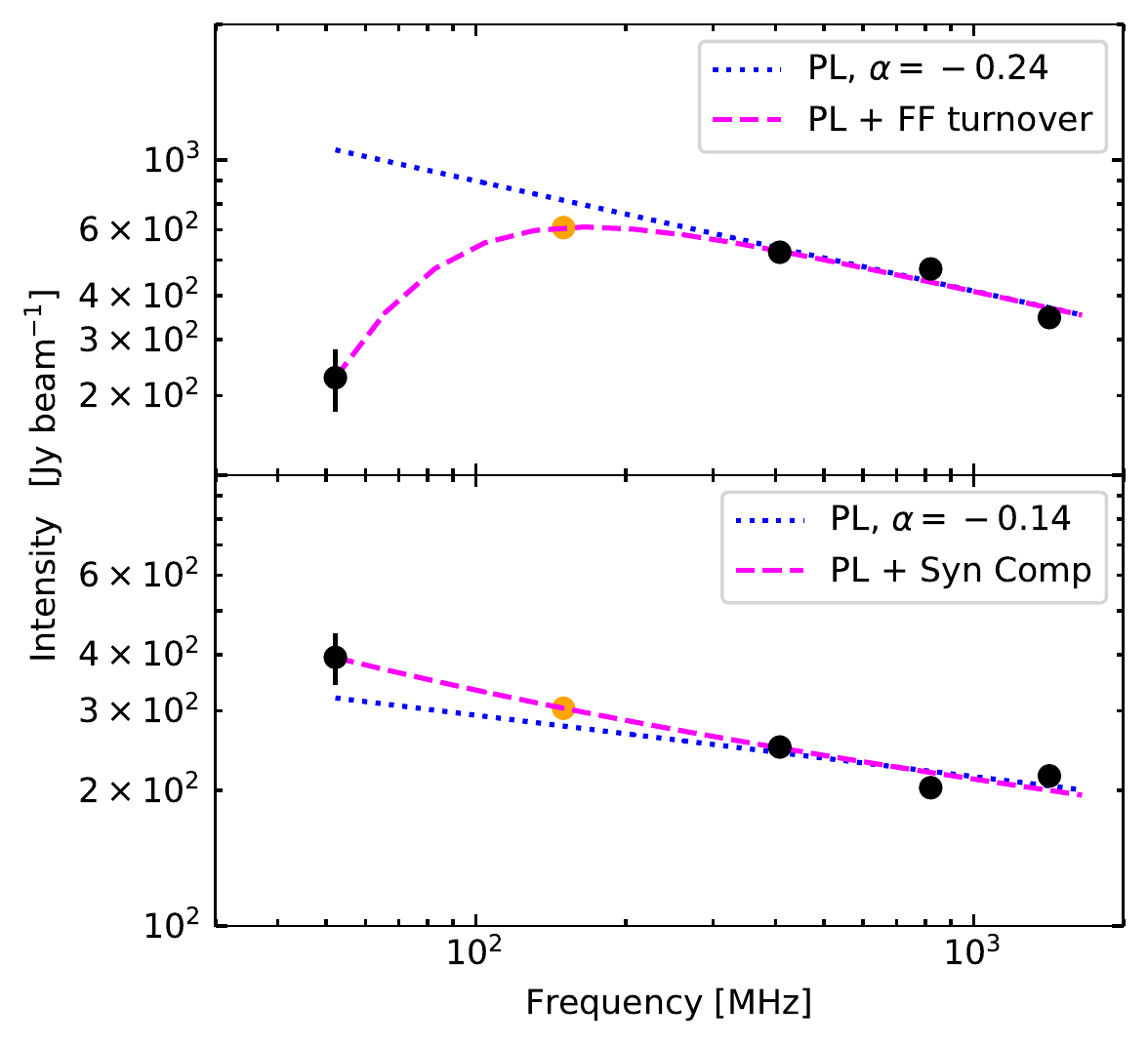}
    \caption{Examples of the multi-frequency fits (done on a pixel by pixel basis) to frequency-interpolate the ``short-spacing'' map -- a low resolution image with sensitivity to large angular scales -- at 148~MHz. The black data points represent the survey data used in the fit. The orange data point represents the interpolated value at 148~MHz. The dotted blue line represents the power-law fit to the high-frequency data points at 408, 820, and 1420 MHz. The dashed magenta line represents the final SED fit -- where either a free-free turnover (\textit{top}) is fit (Equation~\ref{eq:fit_ff}) or a synchrotron component (\textit{bottom}) is added (Equation~\ref{eq:fit_s}) to match the low-frequency point at 52 MHz.}
    \label{fig:ss_fits}
\end{figure}

We construct a short-spacing map at 148 MHz and with 72\arcmin\ resolution that is interpolated from multi-frequency fits across 52~MHz, 408~MHz, 820~MHz, and 1420~MHz. To ensure that $uv$ coverage of the short-spacing map overlaps with the $uv$ coverage of the LOFAR observations, we targeted survey data with resolutions of approximately $\sim$1\degree\ or less and which covered galactic latitudes of $|b| \lesssim 7$\degree.

We compiled the following data. We make use of 52.224~MHz survey data obtained with the Owens Valley Radio Observatory Long Wavelength Array \citep[OVRO-LWA;][]{Eastwood2018}. It has a native resolution of 16.2\amin~$\times$~15.0\amin, and inspection of the data in this region shows that the flux scale is calibrated to within $\sim$20\% as compared to lower resolution surveys. We use the 408~MHz multi-instrument survey of \cite{Haslam1982} that has been destriped (to $<$1~K) of large-scale striations \citep{Remazeilles2015} and has a native resolution of 51\amin \citep{Haslam1974}. The 820~MHz survey with the Dwingeloo Telescope \citep{Berkhuijsen1972} has a native resolution of 72\arcmin, limiting the resolution of our frequency-interpolated map as it has the coarsest resolution of the data used. Lastly, the data used from the Stockert 25~m survey at 1420~MHz \citep{Reich1982} has a native resolution of 35\arcmin. 

The procedure we employ first involves creating an image cutout of 12\degree~$\times$~12\degree\ centered on DR~21 from each survey. Then we smooth the cutouts to the common resolution of 72\arcmin\ and re-grid the images using CASA \citep{McMullin2007, Emonts2019} to a common pixel grid and pixel size of 4.7\amin. Next we convert the intensity scales from temperature brightness units to \Jyb, as the imager we use, \wsc\ \citep{Offringa2014, Offringa2017a}, currently only accepts Jy units. Pixel by pixel, we fit a power-law\footnote{$S(\nu) = S_0 (\nu/\nu_0)^{\alpha}$} to the ``high-frequency'' data points at 408, 820 and 1420~MHz, as previous studies have found this region largely consists of thermal, free-free emission down to 408~MHz \citep{Landecker1984, Xu2013}. Then, to the 52~MHz data point, we fit for either a free-free turnover or a synchrotron component. We do this by extrapolating the high-frequency fit to 52 MHz. If the flux density at 52~MHz is \textit{less} than the extrapolated value of the fit at that frequency, we fit for a free-free turnover,
\begin{equation}
    S(\nu) = \frac{ S_{0,{\rm ff}} }{ \tau_0 }  \left(\frac{\nu}{\nu_0} \right)^2 \left(1 - \exp \left[ -\tau_0 \left(\frac{ \nu }{ \nu_0 } \right)^{\alpha - 2} \right] \right)
\label{eq:fit_ff}
\end{equation}
where $\alpha$ is the spectral index fit to high-frequency data, and $\nu_0 = 148$~MHz such that $S_{0,{\rm ff}}$ is the free-free component of the flux density at $\nu_0$ in the optically thin (unabsorbed) regime and $\tau_0$, the only free parameter in this fit, is the optical depth at $\nu_0$. Otherwise, if the flux density at 52~MHz is \textit{greater} than the extrapolated value of the power-law fit, we add-in a synchrotron component with spectral index $-0.7$ and fit,
\begin{equation}
    S(\nu) =  S_{0,{\rm ff}} \left( \frac{\nu}{\nu_0}\right)^{\alpha} + S_{0,{\rm s}} \left(\frac{\nu}{\nu_0}\right)^{-0.7}
\label{eq:fit_s}
\end{equation}
where $S_{0,s}$ is the synchrotron component of the flux density at $\nu_0$ and the only free parameter in this fit. Synchrotron emission is present from supernova remnants and/or from a diffuse Galactic component. Examples of pixel locations fit with a free-free turnover and with a synchrotron component are shown in Figure~\ref{fig:ss_fits}. Although frequency-interpolation is performed at each pixel independently, point-to-point scatter is not introduced during this process. This is due to the coherence in intensity of neighboring pixels which have been smoothed over by a Gaussian beam that is 235 square pixels in area.

The short-spacing map frequency-interpolated to 148~MHz is shown in Figure~\ref{fig:imaging}\textit{center} for the region of sky falling within the 25\% power of the LOFAR primary beam of the observation. We compare the frequency-interpolated short-spacing map at 148~MHz to an all-sky map at 150 MHz with 5\degree\ resolution \citep{Landecker1970}. The survey data in this region of the sky were originally observed at 178 MHz with 5\degree~$\times$~1.25\degree\ resolution  \citep{Turtle1962} and then scaled to 150~MHz with a spectral index\footnote{$T_b \propto \nu^{\beta}$} of $\beta = -2.6$. Since, the survey data have been re-calibrated to attain an 8~K zero level calibration and a scaling error of 1\% \citep{Patra2015}. We take a cutout from the survey data which covers our region of interest and convert the intensity units from temperature brightness to \Jyb. We convolve our short-spacing map to 5\degree\ resolution and re-grid it to match the 0.23\degree\ pixel scale. Taking the intensity ratio of the 150 MHz survey image over the convolved short-spacing map at 148~MHz results in intensity ratios that have a median value of 1.25. This is within sensible agreement as a 25\% error falls within the uncertainty of our survey data.

%%%%%%%%%%%%%%%%%%%%%%%%%%%%%%%%%%%%%%%%
\subsubsection{Imaging}
\label{sssec:imaging}

\begin{figure*}
    \centering
    \includegraphics[width=0.98\textwidth]{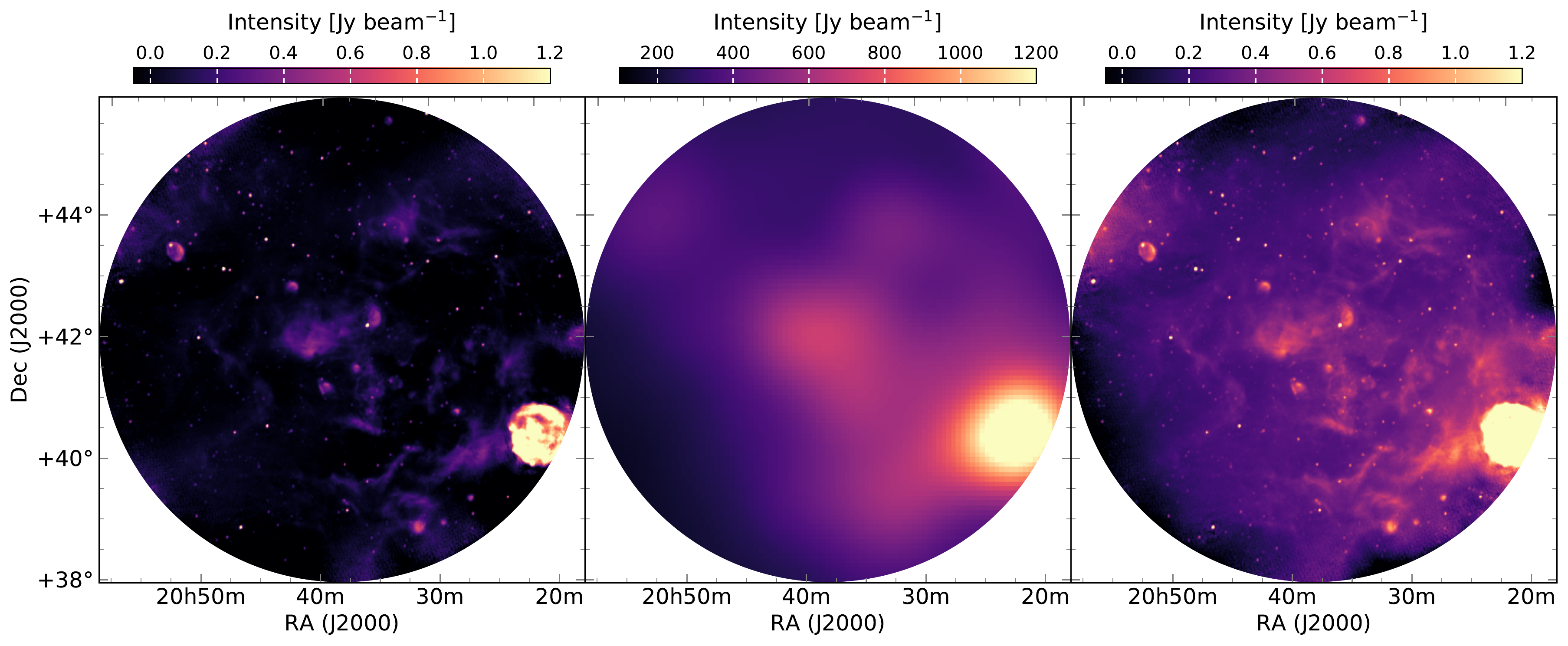}
    \caption{Imaging LOFAR 148~MHz observations of the Cygnus X region. The field of view displayed corresponds to 25\% power of the primary beam. \textit{Left:} Deconvolved LOFAR image without short-spacing information. While these data are sensitive to emission on angular scales as large as 96\arcmin, negative (un-physical) emission is present induced by large scale emission in the Galactic plane. \textit{Center:} The frequency-interpolated short-spacing map at 148~MHz (see Section~\ref{sssec:ss}) and 72\arcmin\ resolution. \textit{Right:} LOFAR data imaged with the short-spacing map as an initial model (see Section~\ref{sssec:imaging}).}
    \label{fig:imaging}
\end{figure*}

The final step in processing the LOFAR observations is imaging the data. We use the frequency-interpolated short-spacing map (Section~\ref{sssec:ss}) as a starting model during imaging in order to deconvolve the interferometric data with zero-level and total flux density information.

To prepare a template for the short-spacing map for \wsc\ \citep{Offringa2014}, we first run \wsc\ with the image weight settings (see below) and 1 iteration so that the pixel grid and synthesized beam size are set. Next we convert the short-spacing map into intensity units of \Jyp according to the synthesized beam size of the \wsc\ image. This is done with the following routine. We re-grid the short-spacing map to match the \wsc\ output pixel grid and size of 18\asec. Then we scale its intensity by the ratio of the beam areas, $(A_{\wsc} / A_{ss})$, to convert units to \Jyb with respect to the \wsc\ output synthesized beam. Next we convert the intensity to units of \Jyp\ by dividing by the beam area in units of pixels.  At this point the short-spacing model is prepared for \wsc\ in true-sky flux density units. \wsc\ requires that we input two model images: one which is attenuated by the primary beam and one which is not. Therefore, we create a copy of our short-spacing model image, but we also apply the primary beam response. We set the primary beam model to zero outside of the first null as we do not image beyond that. With these steps\footnote{We also verify the steps of our template preparation procedure with the CASA preparation procedure outlined by J. Kauffmann at \url{https://sites.google.com/site/jenskauffmann/research-notes/adding-zero-spa} .}, the short-spacing map is prepared to be input to \wsc.

We run \wsc\ using the \texttt{-predict} option to predict the short-spacing model image into the MODEL column of the measurement set. We then run \wsc\ a second time, now with the \texttt{-continue} option to deconvolve the interferometric data. We image out to the first null of the primary beam over an area of 10\degree~$\times$~10\degree. We use Briggs image weighting with a robust parameter of $-0.5$ and a circular Gaussian taper of 1.5\arcmin, effectively smoothing out the long baselines. We CLEAN with multi-scale cleaning \citep{Offringa2017a} on pixel scales of [0,18,36,72,144] and to a threshold of 4$\sigma$ where $\sigma$ is scale dependent and internally calculated. We verified that color corrections do not need to be applied to LOFAR sub-bands, as is standard with LOFAR HBA continuum observations making using of \wsc\ for imaging, in order to obtain accurate monochromatic flux densities. 

In Figure~\ref{fig:imaging}\textit{right}, we show the output short-spacing corrected image. In this paper, we analyze a 4\degree~$\times$~4\degree\ region covered by the LOFAR observations, which approximately encompasses the full-width half power of the primary beam. We smoothed the image to a circular beam of FWHM of 2\arcmin. The final image has a noise of 5~\mjyb\ and effective frequency of 148.05~MHz. 

For comparison, we also image the LOFAR data without the short-spacing map but using the same imaging parameters and show the results in Figure~\ref{fig:imaging}\textit{left}. Additionally, we checked our imaging results by convolving the final image to the 72\arcmin\ resolution of the initial short-spacing map and verified that the total integrated fluxes are consistent. Lastly, we compare our imaging results by ``feathering'' \citep{Stanimirovic2002, Cotton2017} the short-spacing map with the LOFAR data imaged without short-spacings. Through feathering, images are combined in the domain of their Fourier transforms by a weighted average in order to extract the most appropriate spatial frequencies from each image. Taking the intensity ratio of the \casa\texttt{feather} imaged divided by the \wsc\ created image, we find a mean and median value of 0.98 computed over the 4\degree~$\times$~4\degree\ region.

%%%%%%%%%%%%%%%%%%%%%%%%%%%%%%%%%%%%%%%%
\subsection{Ancillary Data}
\label{ssec:ancillarydat}

We compare the LOFAR data with observations compiled through the Canadian Galactic Plane Survey \citep[CGPS;][]{Taylor2003} at 1420~MHz. These data were observed with the Synthesis Telescope at the Dominion Radio Astrophysical Observatory with short-spacing corrections using the Effelsberg Galactic Plane Survey \citep{Reich1990} and the Stockert 25~m survey \citep{Reich1982}. Since the effective resolution varies across the mosaic data products -- the synthesized beam is declination dependent -- we processed the mosaics to attain a common resolution of 2\amin. Then we stitch the smoothed mosaics together. The standard deviation in a relatively low emission region of the image is $\sigma = 0.03$~K (0.7~\mjyb).

%%%%%%%%%%%%%%%%%%%%%%%%%%%%%%%%%%%%%%%%
\section{Continuum Emission}
\label{sec:continuum}

\begin{figure}
    \centering
    \includegraphics[width=0.45\textwidth]{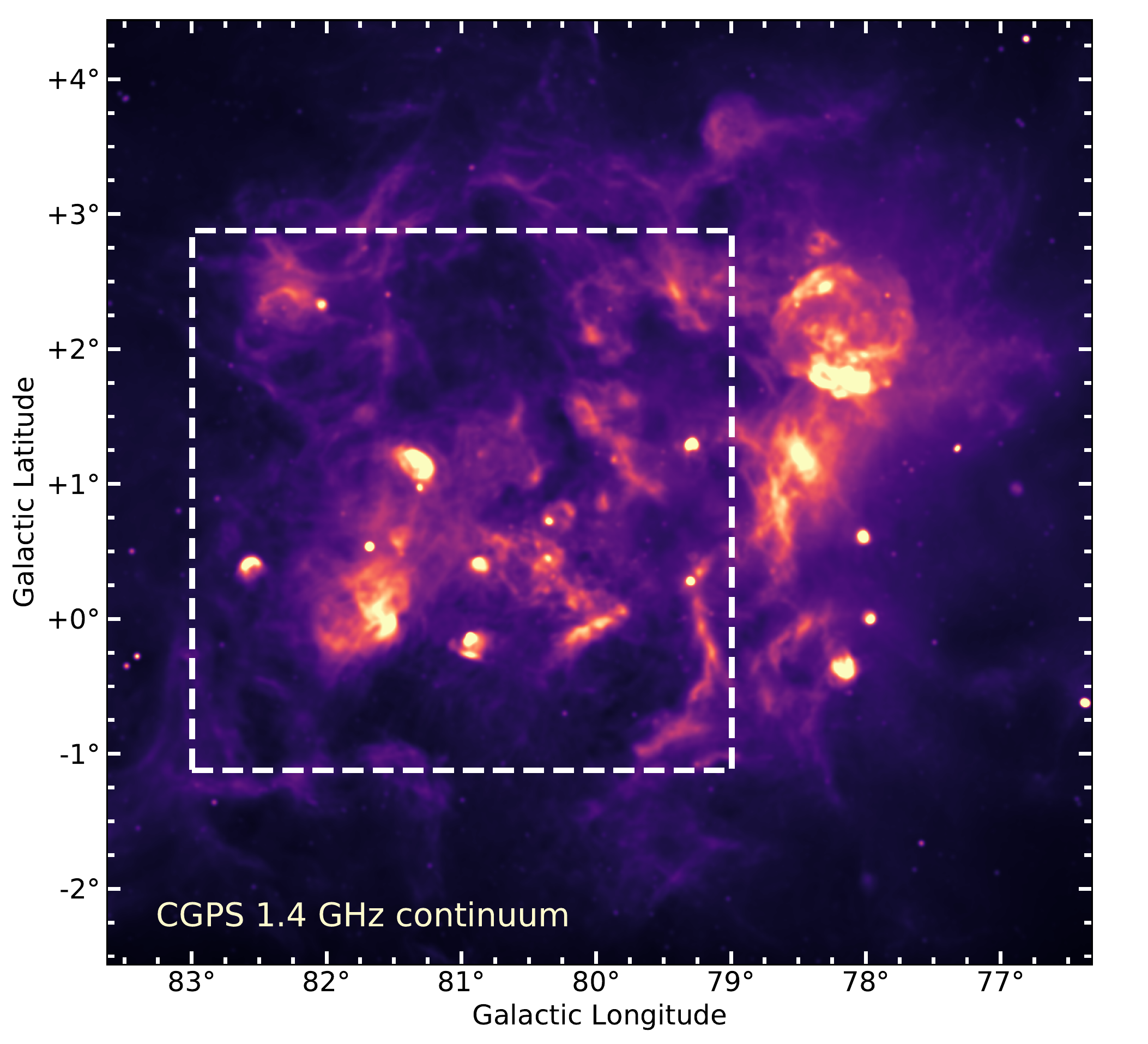}
    \caption{The Cygnus X star-forming region shown in CGPS 1.4~GHz continuum intensity \citep{Taylor2003}. The white dashed lines encompass the region of interest (4\degree~$\times$~4\degree, or  $\sim$100~pc~$\times$~100~pc) that we analyze with LOFAR 148 MHz observations.
    }
    \label{fig:pointingloc}
\end{figure}

\begin{figure*}
    \centering
    \includegraphics[width=0.8\textwidth]{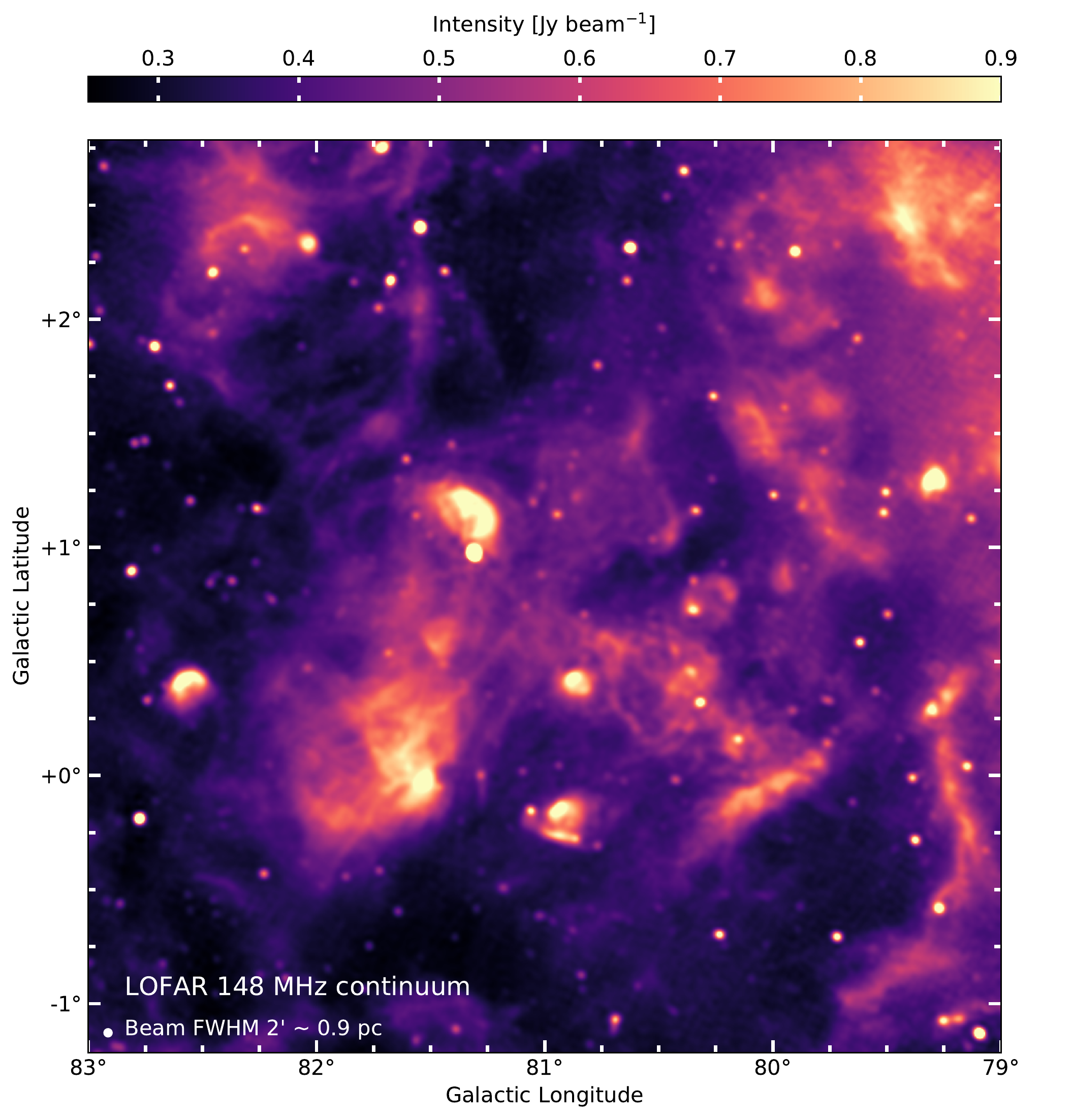}
    \caption{LOFAR 148~MHz continuum emission of the 4\degree~$\times$~4\degree\ ($\sim$100~pc~$\times$~100~pc) area in the Cygnus X region that we analyze at 2\amin\ (0.9~pc) resolution. The emission in this region is primarily thermal, free-free emission from low-density photoionized gas. A number of extra-galactic radio sources are present as bright, point sources. The noise away from bright emission is $\sigma = 5$~\mjyb. }
    \label{fig:fullcont}
\end{figure*}

The Cygnus X star-forming complex encompasses an area of more than 30~square~degrees as shown in Figure~\ref{fig:pointingloc}. The region of interest analyzed with our LOFAR pointing, 4\degree~$\times$~4\degree\ (100~pc~$\times$~100~pc) is depicted with white dashed lines. 

Figure~\ref{fig:fullcont} shows the continuum emission observed at 148~MHz with LOFAR and which is corrected for missing short-baseline information (Section~\ref{ssec:lofarobs}). The 148~MHz emission in this region primarily traces thermal free-free emission from ionized gas with low density and small emission measure \citep[see Section~\ref{sec:ff} and e.g.,][]{Wendker1991, Xu2013}. Thermal radiation referred to as ``free-free'' emission is  bremsstrahlung radiation emitted as the paths of free electrons are deflected in the presence of free ions.

The morphology of the 148~MHz emission includes extended (on degrees scales), resolved regions of photoionized gas in the vicinity of massive stars and star clusters. Filamentary structure is a prominent feature in bright emission regions and is also present, perhaps more fractal-like in faint emission regions. Shell-like regions also appear. An additional diffuse component surrounds much of the extended structures. Extra-galactic radio galaxies with bright synchrotron emission at these frequencies appear as point-like sources at 2\arcmin\ resolution. Analyzing the spectral energy distribution (SED; see Section~\ref{sec:ff}) of unresolved objects is necessary to pull-out synchrotron dominated sources from the regions of active star-formation embedded within a dense medium.  

\begin{figure*}
    \centering    
    \includegraphics[width=0.48\textwidth]{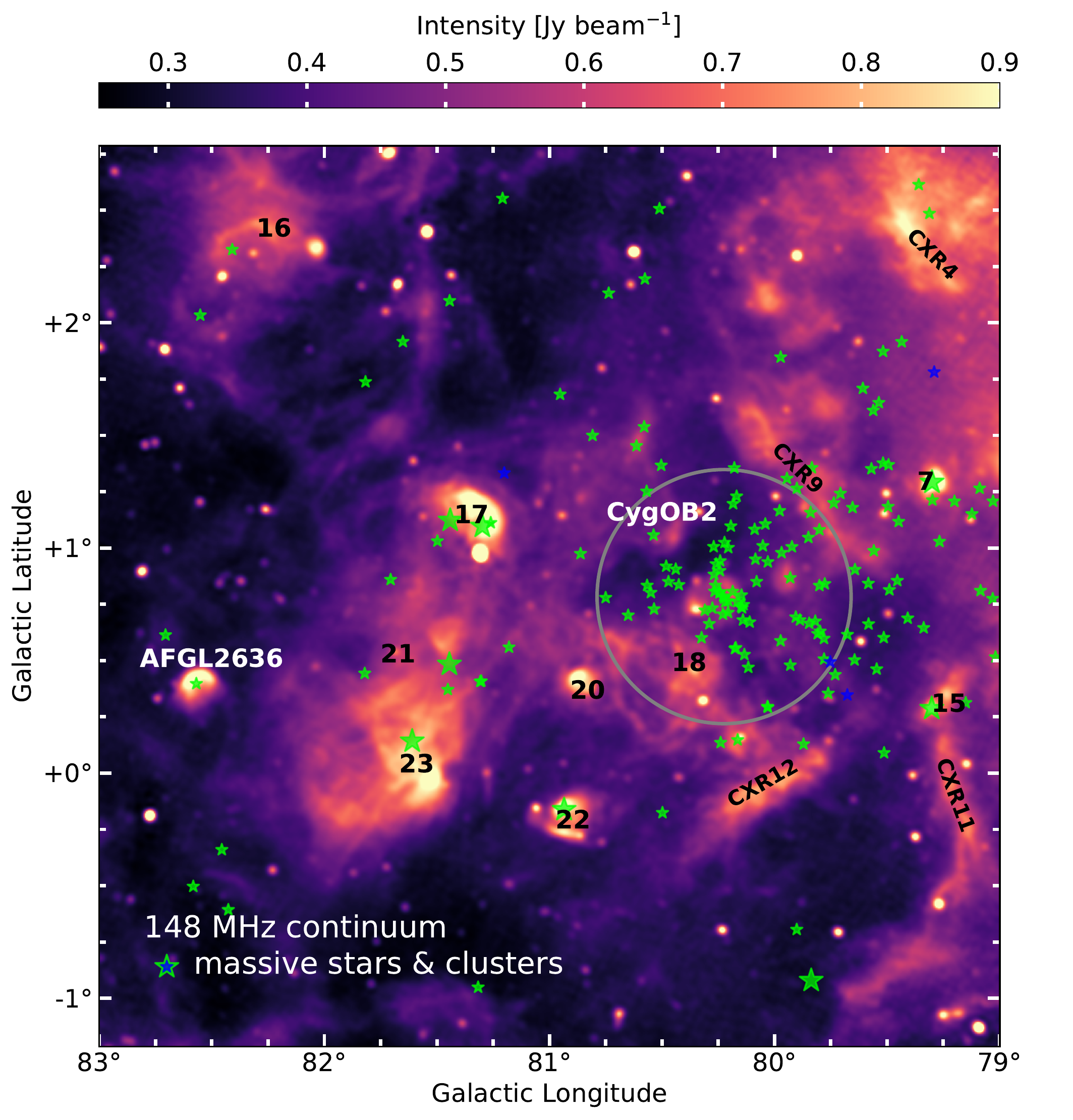}
    \includegraphics[width=0.48\textwidth]{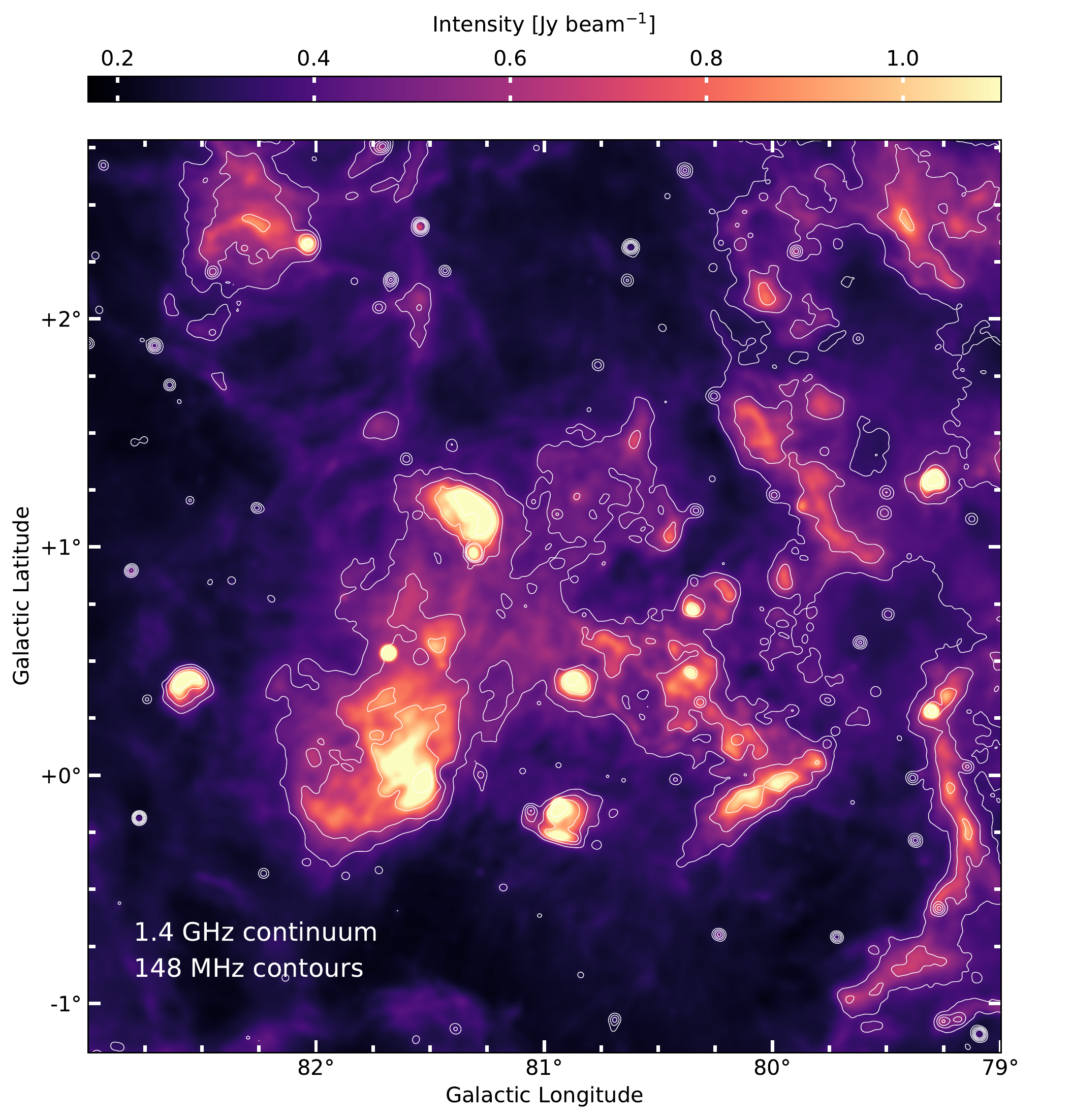}
    \includegraphics[width=0.48\textwidth]{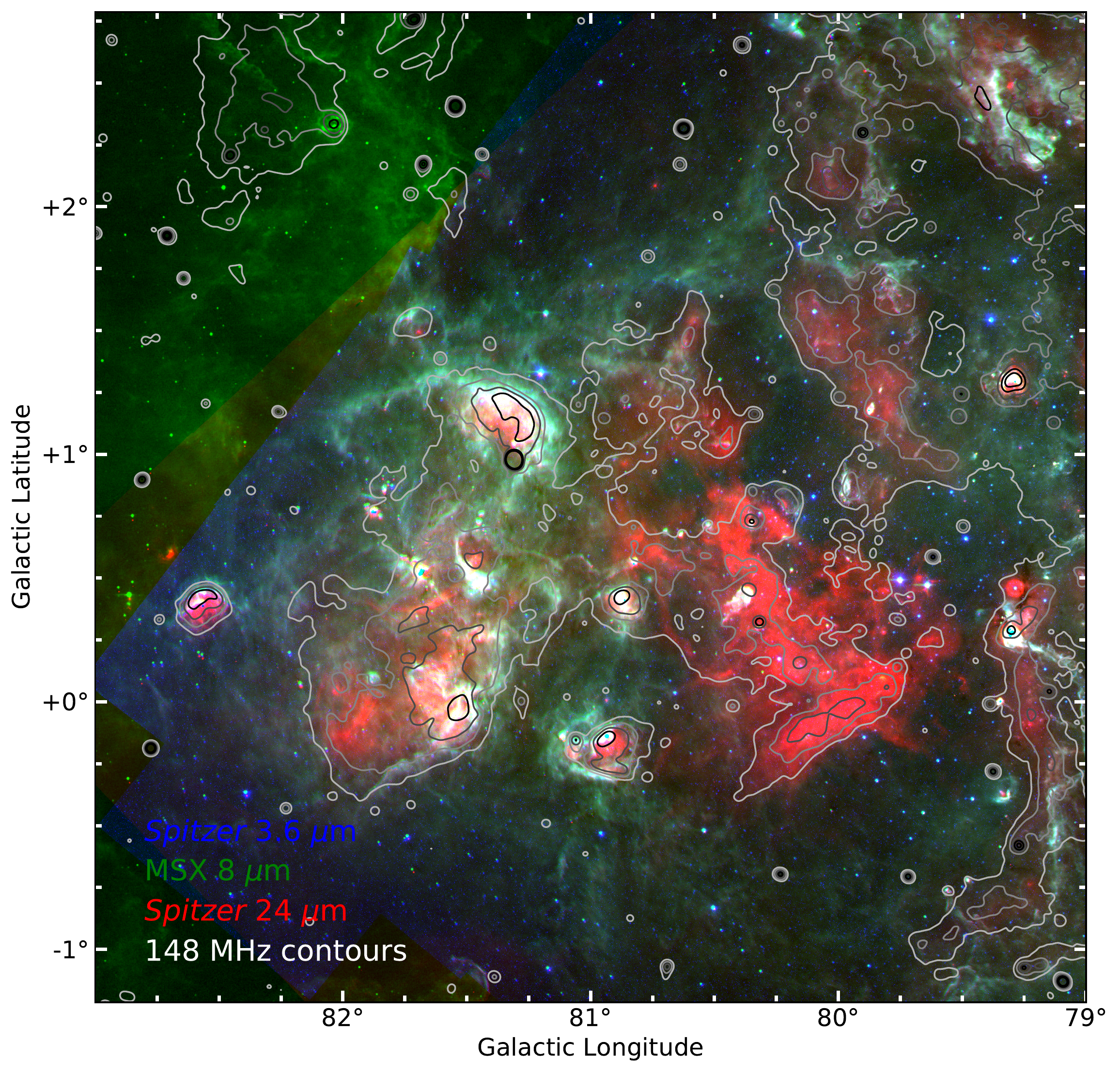}
    \includegraphics[width=0.48\textwidth]{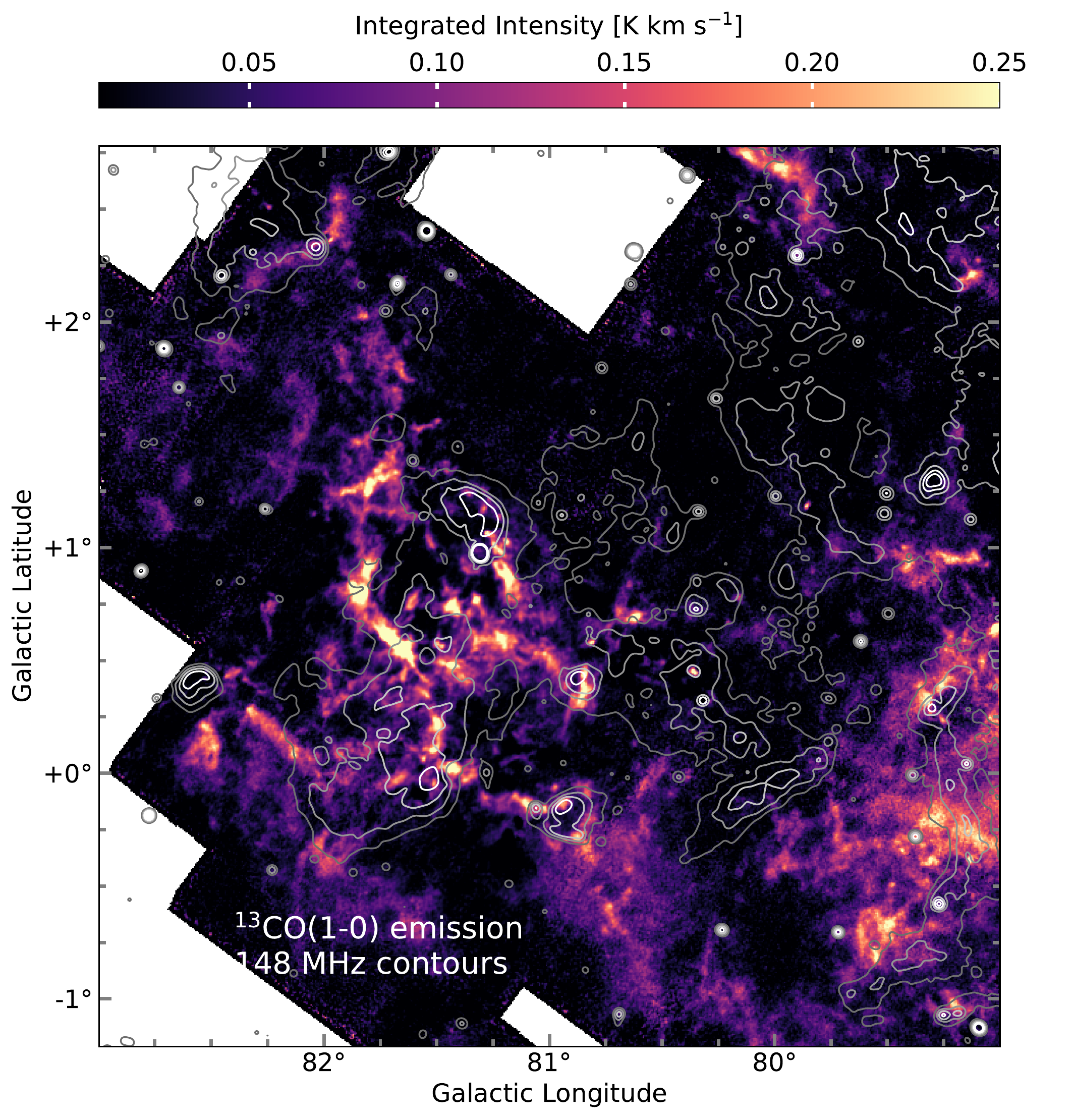}
    \caption{ 
    \textit{Top Left:} The location of massive stars and star clusters are shown overlaid on LOFAR 148~MHz continuum emission. Small green stars represent massive OB stars of the Cyg OB2 association and within the field \citep{Berlanas2018}, and small blue stars mark the locations of supergiants \citep{Comeron2020}. The gray circle represents the core of Cyg~OB2 as identified by \citet{Wright2015}. Large green stars mark the location of (open) clusters in the region \citep{LeDuigou2002}. Numbers in black text denote the position of radio continuum sources as identified by \citet{Downes1966}, and filamentary structures identified by \citet{Wendker1991} are designated with their CXR number. \\
    \textit{Top Right:} 1.4~GHz continuum emission observed through the CGPS \citep{Taylor2003} and smoothed to a common resolution of 2\amin. Contours from the LOFAR 148~MHz continuum are overlaid at $[10^{-0.6}, 10^{-0.55}, 10^{-0.5} ... 10^0]$~\Jyb.
    The morphology in the majority of extended features is consistent with thermal continuum from low-density, ionized gas and where relative differences in the intensity of the emission indicate free-free optical depth effects. \\
    \textit{Bottom Left:} MIR emission. In blue, \textit{Spitzer} 3.6~$\mu$m tracing the stellar population. In green, MSX 8~$\mu$m emission primarily from PAHs heated by UV radiation from massive stars tracing the PDR surfaces. In red, \textit{Spitzer} 24~$\mu$m emission of warm dust, corresponding well with diffuse ionized gas observed at 148~MHz. \\
    \textit{Bottom Right:} $^{13}$CO (1-0) integrated intensity observed with FCRAO \citep{Schneider2007} tracing the bulk of the molecular gas with densities $\gtrsim 3000$~\cmc. 
    \label{fig:comparecont}}
\end{figure*}

The locations of massive stars and star clusters in this region are shown in the top left panel of Figure~\ref{fig:comparecont}. We label the radio continuum features which were identified in \cite{Downes1966} via 5~GHz observations at 10.8\amin\ resolution -- DR 7, 10, 11, 16, 18, 19, 20, 21, 22, 23. Only DR~7 is not associated with the Cygnus X region; instead it is located more than 3.3~kpc distant in the Perseus Arm \citep{Piepenbrink1988}. The Cygnus X region has been extensively surveyed for massive stars and clusters \citep{LeDuigou2002, Comeron2012, Wright2015, Berlanas2018, Comeron2020}. The stars which are members of the Cyg OB2 association are no longer individually embedded within dense ionized gas, while the stars in a number of the DR sources are in the process of dispersing their dense gas.

The 1.4~GHz continuum emission from the CGPS, also shown in Figure~\ref{fig:comparecont}, is strikingly similar to the LOFAR 148~MHz continuum emission. The morphologies and relative intensities of the extended, resolved emission (including filamentary structure) are comparable. The low-level, diffuse emission surrounding extended structures is relatively brighter at lower frequencies. Similarities attest to the primarily thermal nature of the emission \citep{Xu2013, Wendker1991}. Differences in relative intensities arise in compact sources and are attributed to two effects. One, compact regions which appear relatively fainter at 148~MHz have high emission measures ($EM > 10^6$~\emuni) and large optical depths $(\tau_{\mathrm{ff}} > 10)$. Two, at 1.4~GHz the synchrotron emission from extra-galactic sources is faint compared with the thermal component from galactic regions.

Mid-infrared (MIR) emission at 3.6, 8.0, and 24 $\mu$m is compiled as an RGB image in Figure~\ref{fig:comparecont}. Emission at 3.6~$\mu$m in blue and observed with the {\it Spitzer} Space Telescope\footnote{{\it Spitzer} data was acquired from the archive at \url{https://irsa.ipac.caltech.edu/data/SPITZER/Cygnus-X/} .} shows the stellar population of young massive stars as point-like sources of emission \citep{Beerer2010}. 8~$\mu$m emission in green observed with the Midcourse Space Experiment \citep[MSX;][]{Price2001} 
and described by \citet{Schneider2006} especially with respect to CO, mainly traces UV heated polycyclic aromatic hydrocarbons (PAHs) in photodissociation regions (PDRs), thereby emphasizing interfaces between molecular clouds and intense UV fields from young massive stars. {\it Spitzer} 24~$\mu$m emission in red is predominantly produced by thermal emission from warm dust co-spatial with photoioinized gas \citep{Churchwell2006, Calzetti2007, Salgado2012}. Emission that is bright primarily at 24~$\mu$m (red) corresponds well with the ionized gas traced by thermal, low-frequency radio emission. These two tracers correspond well in \Hii\ regions (partially) surrounded by PDR envelopes as well as for ionized regions without PDRs at their edges.

Molecular cloud emission as traced by $^{13}$CO(1-0) \citep{Schneider2007} shows the cold gaseous reservoir in Figure~\ref{fig:comparecont}. $^{13}$CO is largely optically thin (e.g., unlike $^{12}$CO) in the region and traces gas with relatively low densities of $\gtrsim3000$~\cmc. The full Cygnus X complex (see Figure~\ref{fig:pointingloc}) contains $4.7 \times 10^{6}$~\Msun\ of molecular gas \citep{Schneider2006}. High-contrast elongated filaments are observed as well as diffuse emission from higher density clouds. Embedded regions like DR~21, DR~15 and DR~20 are sites of active star-formation coincident with radio emission, whereas DR~17, DR~18, DR~22, and DR~23 are located at the edges of or separated from $^{13}$CO(1-0) emission. Several arc or shell-like structures are also visible in $^{13}$CO(1-0).

%%%%%%%%%%%%%%%%%%%%%%%%%%%%%%%%%%%%%%%%
\section{Ionized gas intensity ratios}
\label{sec:ff}

The flux density observed from optically thin, free-free emission is a function of frequency \citep[e.g.,][]{Condon1992,Emig2020b},
\begin{equation}
\begin{split}
    S_{\rm ff}(\nu)  = 550~\mathrm{mJy}
	\left( \frac{ EM_+ }{ 10^4~\mathrm{cm^{-6}~pc} } \right)
	\left(  \frac{T_e}{ 7400~\mathrm{K}} \right)^{-0.323} \\
	\left( \frac{\nu}{ 148~\mathrm{MHz} } \right)^{-0.118}
\label{eq:I_ff}
\end{split}
\end{equation}
and is dependent on the electron temperature, $T_e$, and the emission measure, $EM_+$, of the ionized medium. The emission measure,
\begin{equation}
    EM_+ = \int n_e n_+ {\rm d} \ell
\label{eq:em}
\end{equation}
is defined by the electron density, $n_e$, the ion density, $n_+$, and the pathlength integral of the emitting regions.

The optical depth of free-free emission is given by \citep{Condon1992, Emig2020b},
\begin{equation}
\begin{split}
\tau_{\mathrm{ff}}(\nu) =  0.29 \left( \frac{ EM_+ }{ \mathrm{10^4~cm^{-6}~pc} }\right)\left( \frac{ T_e }{7400~\mathrm{K}} \right)^{-1.323} \\ \left( \frac{\nu}{ 148~\mathrm{MHz}} \right)^{-2.118}.
\label{eq:tau_ff}
\end{split}
\end{equation}
Self-absorbed, optically thick free-free emission follows a frequency dependence of $S(\nu) = S_0 \nu^{2} (1 - \exp(-\tau_{\rm ff}(\nu)))$, as in Equation~\ref{eq:fit_ff}. A power-law radio spectrum that is absorbed by a free-free component (along the line of sight) is given by $S(\nu) = S_0  \nu^{\alpha} \exp(-\tau_{\rm ff}(\nu)))$.

The radio emission arising from the majority of extended, resolved features in the Cygnus X region is consistent with optically thin thermal, free-free emission at frequencies as low as 408~MHz \citep{Wendker1991, Xu2013}.

\begin{figure}
    \centering
    \includegraphics[width=0.5\textwidth]{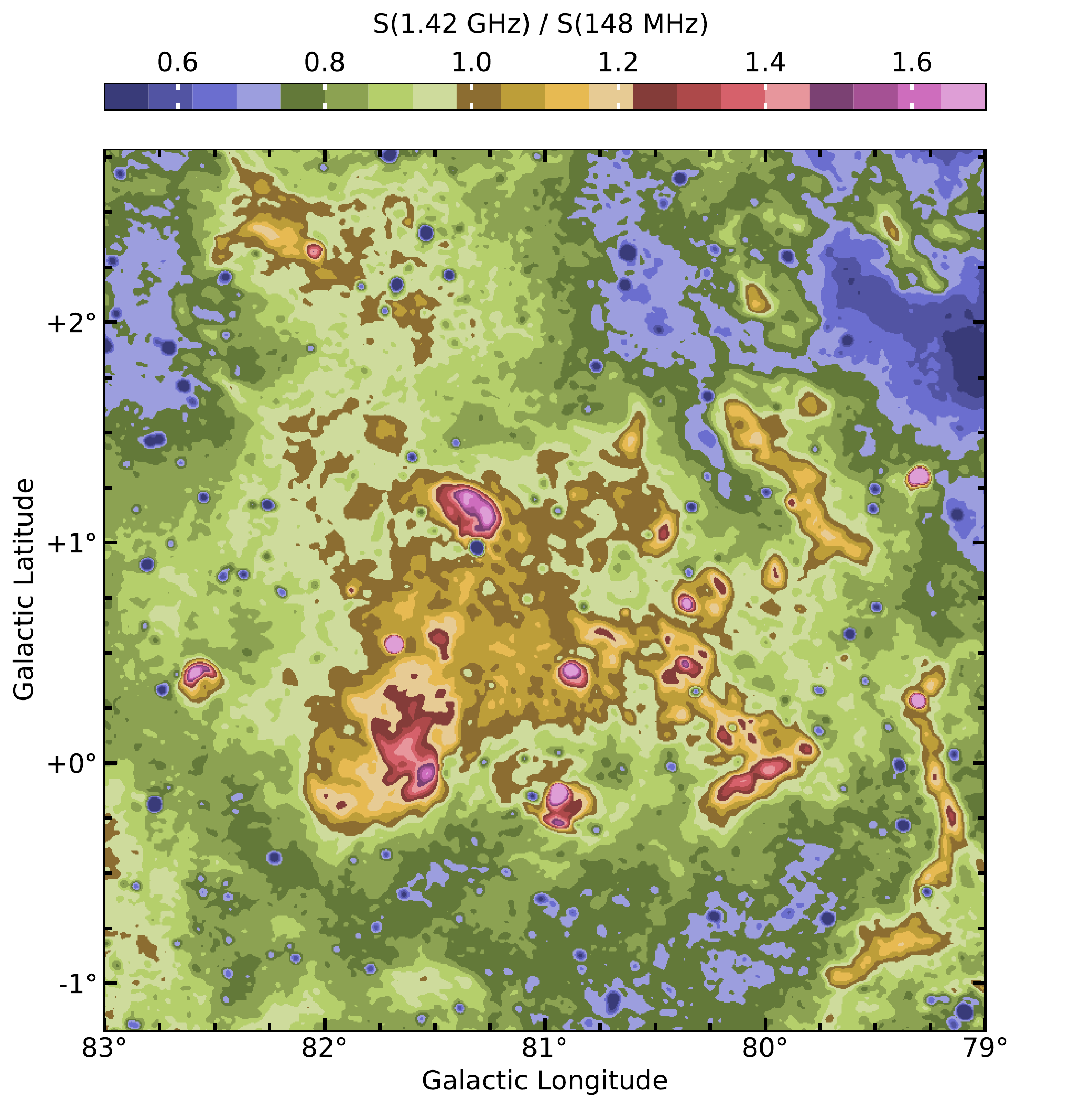}
\caption{ Intensity ratio of continuum emission $S$(1.42~GHz) / $S$(148~MHz). A synchrotron $\alpha = -0.7$ spectrum has an intensity ratio of 0.21 between these two frequencies. An optically-thin free-free spectrum with $\alpha = -0.118$ has a ratio of 0.77, and higher ratios indicate free-free emission that is either self-absorbed or externally absorbed by free-free emission. }
    \label{fig:ratio}
\end{figure}

In Figure~\ref{fig:ratio} we show the intensity ratio of $S$(1.42 GHz) : $S$(148 MHz) where the intensities have units of \Jyb. In order to take this ratio, we regrid the CGPS 1.4 GHz image to the same pixel grid as the LOFAR image. An uncertainty of 25\% in the intensity ratios encompasses systematic uncertainties present in the short-spacing map (see Section~\ref{sssec:ss}). A synchrotron spectrum following a power-law with spectral index $\alpha = -0.7$ has an intensity ratio of 0.21 between these two frequencies. Point-like sources with the lowest intensity ratios are overwhelmingly background synchrotron-dominated galaxies. An optically thin free-free spectrum with $\alpha = -0.118$ has a ratio of 0.77, and higher ratio values indicate free-free emission that is either self-absorbed or externally absorbed by another free-free component. These results show that the region is dominated by free-free emission at 148 MHz, in the smaller scale structures as well as in much of the larger scale, diffuse emission. In a large portion of the image, 82\% of pixels, the intensity ratio is $>0.77$ and appears to be dominated by free-free emission showing optical depth effects. The largest ratios are found in the compact sources of DR~7, DR~15, DR~21, and DR~22, for which optical depth effects are already important (e.g., $\tau > 0.1$) at 1.4~GHz \citep{Wendker1991}. 

We also conclude from Figure~\ref{fig:ratio} that two or more free-free components contribute to the observed continuum emission. This includes emission from the named regions and resolved structure, as well as large scale emission. Specifically the emission outside of the resolved structure does not appear to be largely uniform in its intensity and physical properties. 

To estimate its properties and check our flux scale, we determine the internal and external (line of sight) components estimated from the optically thin 1.42 GHz emission towards three sources. Using circular apertures with a 1\arcmin\ radius, we extract flux from regions within DR23, CXR12, and DR17. The total flux densities at 1.42 GHz are measured to be $S_{\rm T}$(1.42 GHz) = 3.3, 2.4, and 3.5 Jy, respectively. We determined external emission to be $S_{\rm e}$(1.42 GHz) = 0.7, 0.6, and 0.9 Jy in an uncontaminated region about 8\arcmin\ away on average from each source. Subtracting the external emission, leaves internal emission to the sources of $S_{\rm i}$(1.42 GHz) = $S_{\rm T}$ - $S_{\rm e}$ = 2.6, 1.8 and 2.6 Jy. At the same locations in the 148 MHz continuum, we measure flux densities of $S_{\rm T}$(148 MHz) = 2.1, 1.7, and 2.2 Jy respectively.

If we were to fit for the free-free optical depth at 148~MHz, by comparing $S_{\rm T}$ at the two frequencies, using Equation~\ref{eq:fit_ff}, and assuming the emission only comes from a single component, the optical depths would be $\tau$(148~MHz) = 1.7, 1.4, and 1.8 for regions in DR23, CXR12, and DR17, respectively. By using these values and combining Equations~\ref{eq:I_ff} and \ref{eq:tau_ff}, we solve for the electron temperature and find 4400, 4100, 4500 K. This is in contrast with a median temperature of $T = 7400$~K derived from radio recombination line observations at 4.8~GHz and 2.6\amin\ resolution \citep{Piepenbrink1988}. In other words, when we only fit one free-free component, the flux appears to be too low at 148 MHz and predicts too large of an optical depth.

Next, we consider two free-free components. One external line-of-sight component, which may have optical depth effects present and thus follows $S_e(\nu) = S_{0, \rm e} \nu^{2} (1 - \exp(-\tau_{\rm e}(\nu)))$, as in Equation~\ref{eq:fit_ff}. A second, internal (to Cygnus X's resolved structure) component which may have optical depth effects present, but in addition, could also be absorbed by optically thick free-free material along the line of sight, and follows $S_{\rm i}(\nu) = S_{0,\rm i} \nu^{2} (1 - \exp(-\tau_{\rm i}(\nu))) \exp(-\tau_{\rm e}(\nu))$. The observed spectrum would be the summation of these two components $S_{\rm T} = S_{\rm e} + S_{\rm i}$. In this scenario, when plugging in the values for $S_{\rm e}$(1.42 GHz) and $S_{\rm i}$(148 MHz) as we defined above and letting $T = 7400 K$, free-free optical depths are calculated to be $\tau_{\rm i}$ = 0.8, 0.5, and 0.8, and $\tau_{\rm e}$ = 0.5, 0.5, 0.5 for the regions DR23, CXR12, and DR17, respectively. And, we recover the expected flux at 148 MHz to within ten percent, $S_{\rm T, predicted}$(148 MHz) = 2.3, 1.8, and 2.3 Jy respectively. This shows that the emission can be modeled as two free-free components and verifies the flux in our observations.

Unfortunately, the two point SED at 2\amin\ resolution does not allow us to fit for multiple continuum components and free-free optical depth effects and perform this analysis in a systematic way for the whole map. We also tested whether additional (sub) images created across the HBA band -- at 120.9, 137.2, 149.4, and 170.7 MHz -- could be used to fit multiple SED components. For bright sources in the field, we do confirm a trend consistent with free-free emission within the LOFAR band. However, in the majority of the map, this subdivision did not provide reliable constraints on the contribution of additional continuum components. Instead in the next section, we focus only on the resolved filamentary like components. We use \filchap\ to subtract out external emission and isolate intrinsic emission which is largely optically thin at 1.4 GHz and can thus be described by Equation~\ref{eq:I_ff}. In the future, LOFAR LBA observations at 30--80 MHz will better constrain SED modeling, especially in the presence of multiple SED components.

%%%%%%%%%%%%%%%%%%%%%%%%%%%%%%%%%%%%%%%%
\section{Analyzing filamentary structure} 
\label{sec:filaments}

Filamentary structure is a prominent feature of the low-density ($n_e \lesssim 100$~cm$^{-3}$) ionized gas in the Cygnus X region, as we point out in Section~\ref{sec:continuum} and which is also seen over a larger area of the Cygnus X region in Figure~\ref{fig:pointingloc}. We are motivated to derive properties of this filament-like emission to investigate how ionized gas and ionizing radiation interact with and shape their environment.
To briefly summarize the contents of this section, Figure~\ref{fig:fil_loc} shows filament spines identified with the \disperse\ algorithm \citep{Sousbie2011}. Figure~\ref{fig:fil_fit} demonstrates the processing and fitting of the radial distribution of a filament profile using \filchap\ \citep{Suri2019}. We plot the distributions of the peak $EM$ fit to the radial profiles, the best-fit widths, and the inferred electron densities in Figure~\ref{fig:fil_hist}. The peak $EM$s of the filament profiles follow a power-law in their number distribution down to the estimated completeness limit. The widths of the filament profiles show a characteristic peak at a median value of 4.3~pc (noting that our beam resolution is 0.9~pc).  The median electron density within filament profiles is $n_e = 35$~\cmc, with densities spanning $n_e = 10 - 400$~\cmc. Figures~\ref{fig:fil_widthmap} and \ref{fig:fil_pw} compile and compare the peak $EM$, width and density.

%%%%%%%%%%%%%%%%%%%%%%%%%%%%%%%%%%%%%%%%
\subsection{Identifying filamentary structure}
\label{ssec:identifyfil}

\begin{figure}
    \centering
    \includegraphics[width=0.45\textwidth]{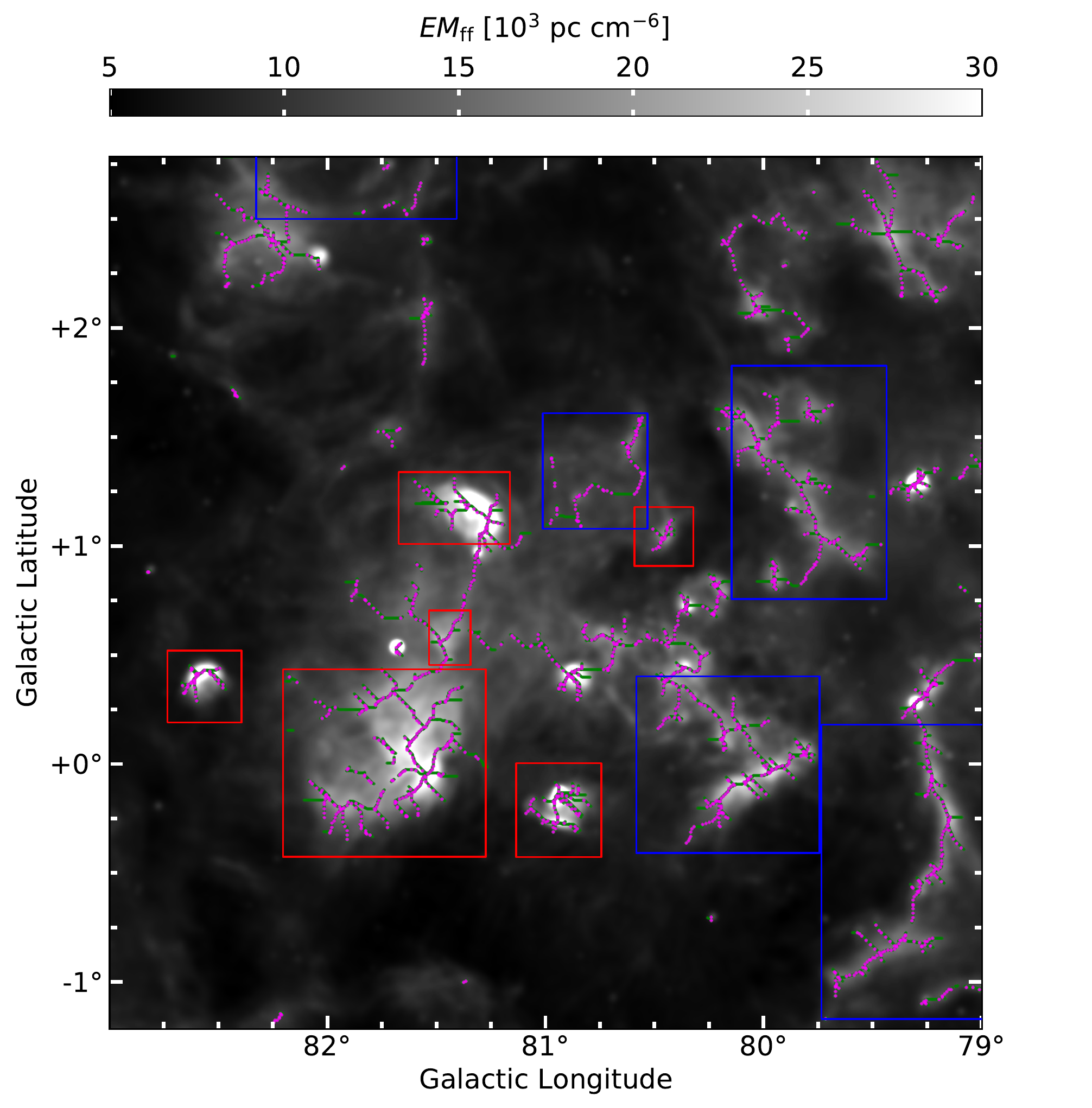}
    \caption{Filaments identified with \disperse\ in green. Purple data points mark the locations along the filament of the average radial profiles which contributed to the analysis. The gray-scale image is the emission measure, computed from the 1.4~GHz continuum intensity assuming optically thin thermal emission, and is the image on which the filaments are identified. Red (Blue) boxes encompass select filaments identified with (out) a known massive star which locally contributes to ionization.}
    \label{fig:fil_loc}
\end{figure}

Using the Discrete Persistent Structures Extractor \citep[\disperse;][]{Sousbie2011}, we identify filamentary structure in the Cygnus X Region as shown in Figure~\ref{fig:fil_loc}. \disperse\ uses discrete Morse theory to derive information on the topology of a given data set. Filaments are identified as the set of arcs joining maxima and saddle points. Persistence theory is used to filter out and identify filaments with significance. We refer the reader to \citet{Sousbie2011} for a detailed description of the procedures. \disperse\ has been widely used to extract filamentary structures both in observational and simulated datasets \citep[e.g.,][ and references therein]{Andre2014}.

We apply the \disperse\ algorithm to a map of the emission measure derived from the CGPS 1.4 GHz intensity using Equation~\ref{eq:I_ff}. 
We filter the filament identification output of \disperse\ using the {\tt -breakdown} option to merge overlapping filament segments and with {\tt -trimBelow} to remove arcs below a robust persistence\footnote{for the specific implementation see \url{http://www2.iap.fr/users/sousbie/web/html/index959e.html?post/definitions}} of 5300~\emuni.  The rms in low intensity regions of the $EM$ map, interpreted as the offset value or background contribution, is 5300~\emuni\ and the standard deviation, interpreted as the noise,  is $\sigma \approx 53$~\emuni. Our input parameters to \disperse\ result in local maximum and saddle point peak $EM$ intensities of $\gtrsim 9500$~\emuni\ being identified. Filaments are output as lists of image coordinates sampled with points at each half pixel shift in direction. 

We select an effective threshold of 9500~\emuni\ because it qualitatively performed well in recovering structures expected by eye. When lowering the threshold, filament segments were found in the brightest regions which did not seem to correlate well with locally bright extended segments of emission. However, a consequence of the high threshold is that we do not include some of the faintest filament-like structures that are clearly discernible by eye.

%%%%%%%%%%%%%%%%%%%%%%%%%%%%%%%%%%%%%%%%
\subsection{Fitting profiles of filamentary structure}
\label{ssec:fitfil}

\begin{figure}
    \centering
    \includegraphics[width=0.45\textwidth]{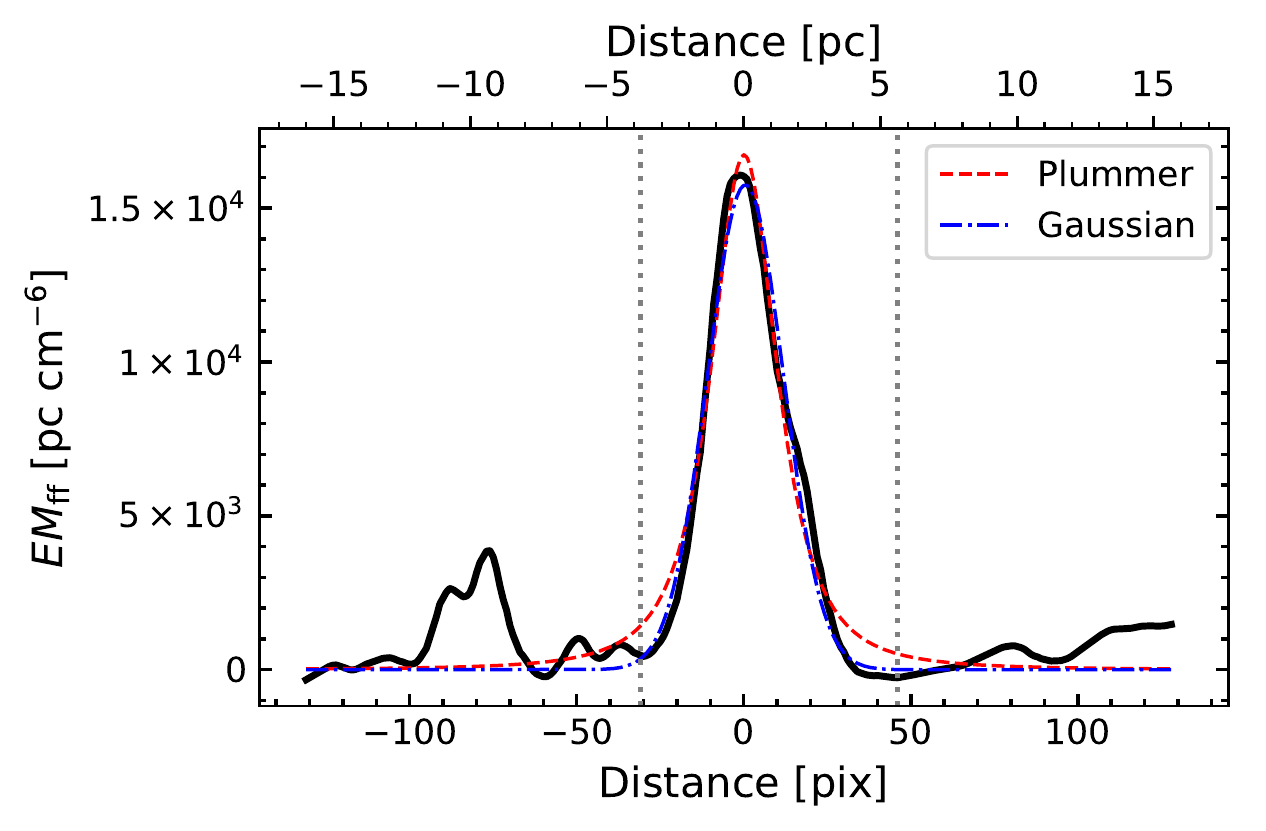}
    \caption{An {\it EM} radial profile of a filament segment, shown with a black solid line. The radial profile has been baseline (and background) subtracted. Negative distances from the spine point to the northeast direction, and positive distances to southwest of the filament. The blue dot-dashed line shows the best fit Gaussian used to compute the filament profile properties. For comparison, the red dashed line shows the best fit Plummer profile with a power-law index of $p=4$.}
    \label{fig:fil_fit}
\end{figure}

We characterize properties of filamentary structure using the python-based Filament Characterization Package \citep[\filchap;][]{Suri2019} which was designed to work together with \disperse. Here we give a brief explanation of the procedures implemented in \filchap\ and refer the reader to \cite{Suri2019} for more detailed descriptions. \filchap\ takes as input a list of filament coordinates. 
At each sample point along the filament, a radial profile perpendicular to the filament spine is extracted. The radial profile extends to $\pm130$ pixels, which we choose so that a baseline estimate includes true background emission at any location in the image.\footnote{We padded the image using the true $EM$ intensity so that filament profiles at edges of the region would be fully sampled.}
The radial profiles extracted at four consecutive sample points of the filament are averaged together. This mean radial profile is baseline subtracted, effectively removing any background emission and baseline gradient. To the mean radial profile, \filchap\ fits three line profiles --- a Gaussian and two Plummer-like functions --- and computes the second moment\footnote{The $n$th moment of a distribution $I$ is given by, $m_n = (N\sigma^n)^{-1} \Sigma_i^N I_i (x_i - \bar{x})^n$, where $\bar{x}$ is the intensity weighted mean position of the profile, $I_i$ is the intensity at position $x_i$, and $\sigma$ is the intensity weighted standard deviation of the profile.} width. Plummer-like functions have been used to describe the column density of a filament with a dense and flat inner portion and a power-law decline at larger radii \citep[e.g.,][]{Arzoumanian2011}. The filament properties that we report are consistent with the four different types of fits and determinations. For simplicity, we use the properties derived from Gaussian fits to represent our results. We demonstrate a radial profile processed and fit with \filchap\ in Figure~\ref{fig:fil_fit}. 

Through the line fitting process, 2141 mean radial profiles are attempted to be constructed. We remove fits which failed or indicated they might be poor fits with Gaussian widths less than 2 pixels, resulting in 1874 filament profiles that we analyze.

%%%%%%%%%%%%%%%%%%%%%%%%%%%%%%%%%%%%%%%%%%%%%%%%%%%%%%%%%%%%
\subsection{Properties of filamentary structure}
\label{ssec:filprops}

\begin{figure*}
    \centering
    \includegraphics[width=0.95\textwidth]{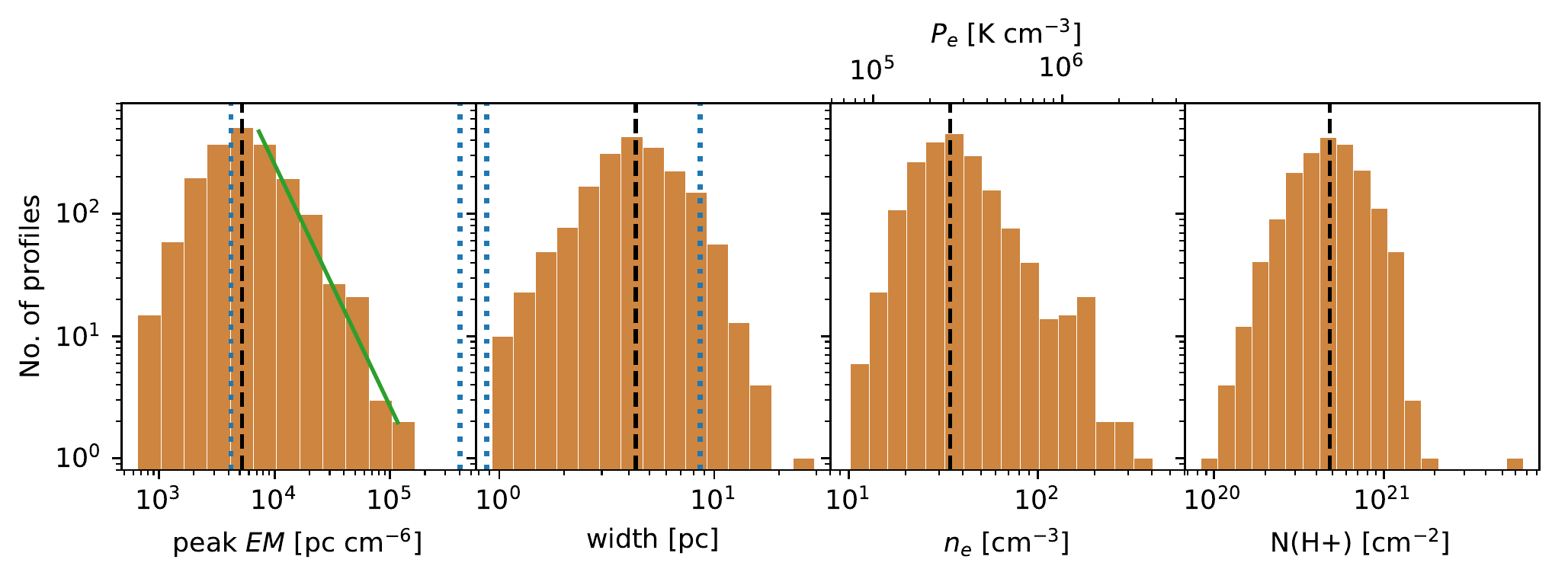}
    \caption{Histograms of the properties of filamentary structure inferred from Gaussian fits to their $EM$ radial profiles. Black dashed lines mark the median values. Blue dotted lines mark the estimated completeness limits in the $EM$ and width plots. 
    {\it Left:} Peak emission measure ($EM$) of the best fit in units of \emuni.  The median $EM$ of the filamentary structure identified in this analysis is 5200~\emuni. The green solid line shows the power-law, $\beta = -2.0 \pm 0.2$, fit to the distribution for $EM > 5200$~\emuni.
    {\it Center left:} Filament profile FWHM in units of pc. The median FWHM of the filamentary structure identified in this analysis is 4.3~pc. The resolution of the beam FWHM (2\amin $\sim$ 0.9~pc) is marked (with a red dotted line) as the lower completeness limit. Widths smaller than the beam resolution result from poor fits.
    {\it Center right:} Average electron density of the filament profiles in units of cm$^{-3}$, computed with the best fit peak $EM$ and width as $n_e = \sqrt{EM / \ell} $ where $\ell$ is the FWHM of the Gaussian profile. The median electron density of the filamentary structure identified in this analysis is 35~cm$^{-3}$. The electron pressure, computed as $P_e / k_{\rm B} \sim 7400~\mathrm{K} \cdot n_e$, is shown along the top axis.
    {\it Right:} The column density of ionized hydrogen, letting $n_e \sim n_{\rm H^+}$.
    }
    \label{fig:fil_hist}
\end{figure*}

%%%%%%%%%%%%%%%%%%%%%%%
\subsubsection{Peak EM of filamentary structure}
\label{sssec:fil_peak}

% peak
\filchap\ determines the peak of the $EM$ radial profile (which has been background and baseline subtracted) from the best fit of a Gaussian profile to 1874 mean radial profiles extracted within filaments. In Figure~\ref{fig:fil_hist}, we plot a histogram of the peak $EM$s.  We find peak {\it EM}s ranging from 500~\emuni\ to $10^5$~\emuni, with a median value of $5.2 \times 10^3$~\emuni.

Regarding the completeness of the distribution, we expect that emission measures in the range $4 \times 10^3 \lesssim EM \lesssim 4 \times 10^5$~\emuni\ are complete. 
From Equation~\ref{eq:tau_ff}, we have that at 1.4 GHz a free-free optical depth of $\tau_{\rm ff}(1.4~\rm{GHz}) = 0.1$ (1) corresponds to $EM = 4 \times 10^5$ $(10^6)$~\emuni. Thus we expect to measure an upper limit of $EM \lesssim 4 \times 10^5$~\emuni. Since the rms in low intensity regions is equal to $5.3 \times 10^3$~\emuni, we expect an upper bound to fully sampled emission measures of $< 3.5 \times 10^5$~\emuni. Furthermore, we applied a cut to select filaments with a peak $EM > 9.5 \times 10^3$~\emuni. Subtracting the ``background'' $EM$, which will be greater than or equal to $5.3 \times 10^3$~\emuni, results in an estimated completeness of $4.2 \times 10^3$~\emuni. Therefore, filamentary structure with a fitted peak $EM$ below $4.2 \times 10^3$~\emuni\ are not fully sampled.

We fit a power-law to the number distribution of the peak $EM$ for bins which we consider to be complete. This results in a best fit power-law index of $\beta = -2.0 \pm 0.2$.

%%%%%%%%%%%%%%%%%%%%%%%
\subsubsection{Widths of filamentary structure}
\label{sssec:fil_width}

In Figure~\ref{fig:fil_hist}, we plot the best fit FWHM of each mean radial profile, converted into physical units assuming the average distance of $d = 1.5$~kpc to the complex. A peak in the width distribution arises at 4.3~pc, which is well separated from the beam resolution of 0.9~pc (2\amin). 
While the \disperse\ algorithm does not prevent large scale features from being identified, there are regions of diffuse emission that could possibly be identified as ``filaments'' at lower resolution. For the $EM$ we probe, we place a conservative upper completeness limit to the widths of filamentary structure of 8.6~pc, twice the median value.

\begin{figure}
    \centering
    \includegraphics[width=0.45\textwidth]{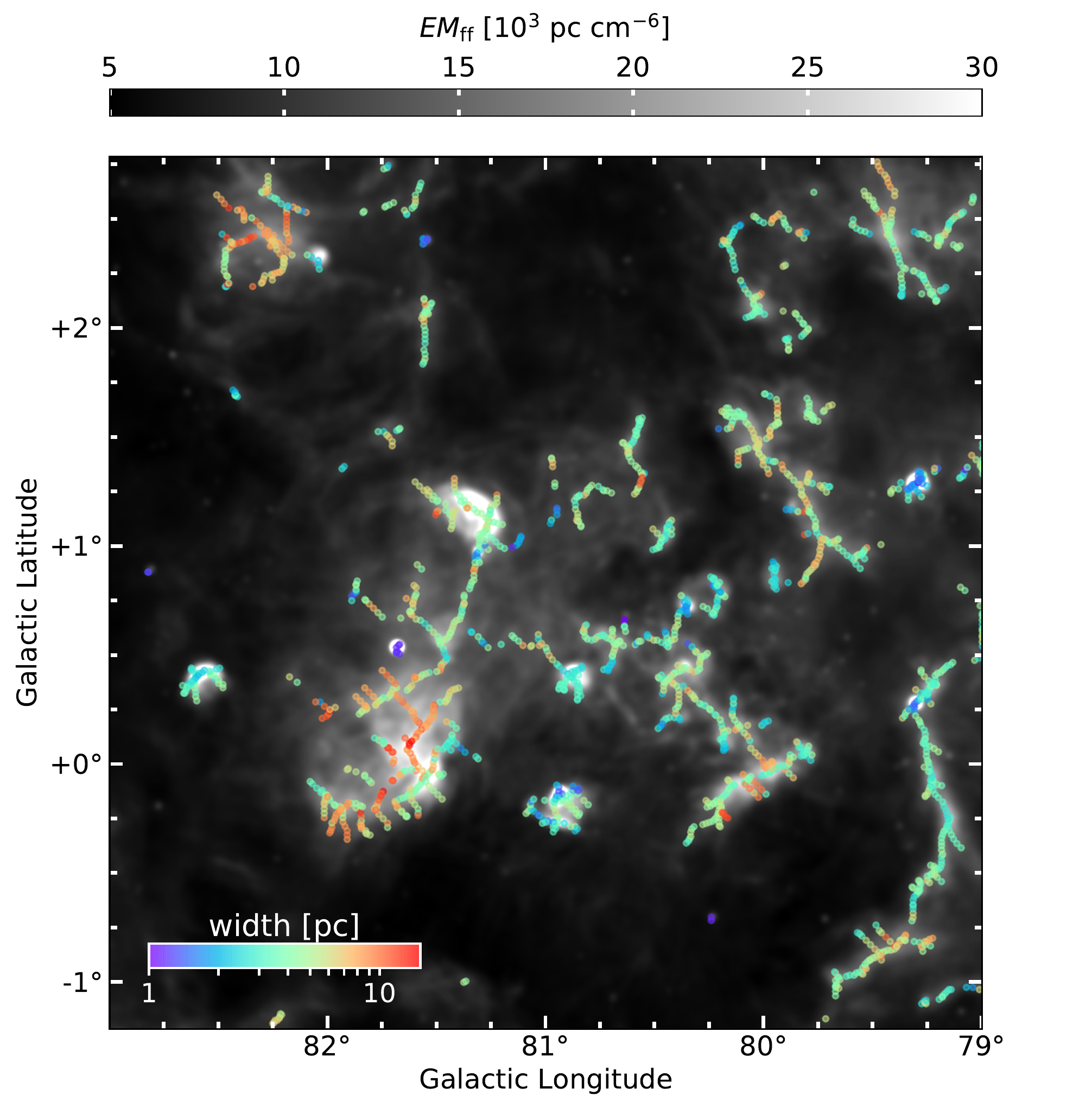}
    \includegraphics[width=0.45\textwidth]{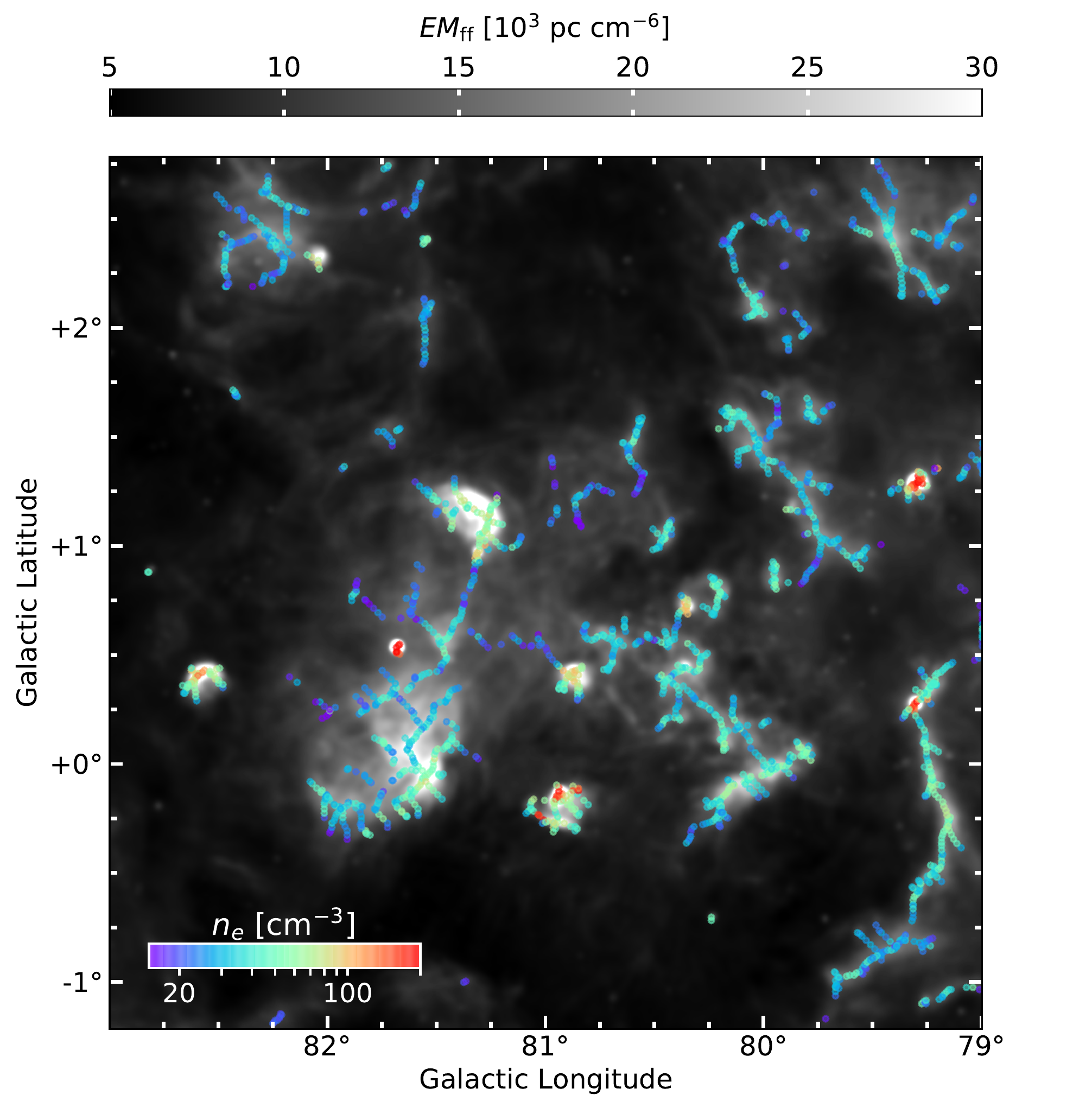}
    \caption{\textit{Top:} Width of filaments' mean radial profiles are represented as colored data points and overplotted on the {\it EM} map. Compact bright regions, tend to have smaller widths. Another notable is the DR~23 region which has wider filamentary structure than average. \textit{Bottom:} The estimated density of each radial profile are represented as colored data points and overplotted on the {\it EM} map. As expected, densities are highest in known star-forming regions.
    }
    \label{fig:fil_widthmap}
\end{figure}

In Figure~\ref{fig:fil_widthmap} we show the widths of filamentary structure with respect to their spatial location on the $EM$ map. Bright compact regions of emission tend to have smaller widths (in blue). Also notable is the DR~23 region which has somewhat wider filamentary structure (in orange, red) than average. While projection effects and variations in distances to the emission features could reasonably influence the apparent widths of the filaments, it is curious that a characteristic value of 4.3~pc is seen throughout the region.

%%%%%%%%%%%%%%%%%%%%%%%
\subsubsection{Electron densities of filamentary structure}
\label{sssec:fil_ne}

We estimate the electron density of the filamentary structure from the best fit peak {\it EM} and FWHM width of the line profile, as $n_e = \sqrt{ EM / {\rm (width)} }$. This calculation assumes that the size of the gas cloud along the line-of-sight is equal to its projected dimension in the plane of the sky. Although this geometry is a good approximation for true filaments, a sheet-like morphology may be a more realistic approximation for the ionized surfaces of molecular clouds; we discuss this in more detail in Section~\ref{sssec:photoevap}. In this scenario, the pathlength along the line-of-sight may be longer, which would imply these densities are upper limits. Perusal of our map indicates that the density would be overestimated by at most a factor of three, and an approximate factor of two in uncertainty of the electron density is reasonable.  Lastly, we acknowledge the gas may be clumpy or show substructure on scales smaller than the resolution of the observations; volume filing factors may be $<1$ for gas of various densities \citep[e.g.,][]{Gao2015}.

In Figure~\ref{fig:fil_hist}, we plot the distribution in electron density of the mean radial profiles. Electron densities ranging from 10~cm$^{-3}$ to 400~cm$^{-3}$ are found, with a median value of 35~cm$^{-3}$. In Figure~\ref{fig:fil_widthmap} we show the filamentary structure densities with respect to their spatial location on the $EM$ map. We note that our census of electron densities may not be complete in the range of values we derive. For example, we find electron densities of $n_e = 20$~\cmc\ in some of the ionized ridges. At a size scale equal to our resolving beam, 0.9~pc, an emission measure of 360~\emuni\ would result, and our analysis would not be sensitive to it.

%%%%%%%%%%%%%%%%%%%%%%%
\subsubsection{Column density of ionized gas}
\label{sssec:fil_coldens}

From the density and width we compute the column density of ionized gas. We assume the gas is completely ionized, with $n_e \sim n_{\rm H\textsc{ii}}$, and compute a column density of $N($\Hii$) = \int n_e {\rm d}\ell = \sqrt{EM \cdot {\rm width}} = EM / n_e $. The ionized hydrogen column densities we derive range from about $(10^{20} - 10^{21})$~cm$^{-2}$ with a median value of $4.8 \times 10^{20}$~cm$^{-2}$. However, given how we derive the column density, we do not expect our census to be complete for this range of column densities.

%%%%%%%%%%%%%%%%%%%%%%%
\subsubsection{Pressure of ionized gas}
\label{sssec:fil_press}

We calculate the electron pressure, as $P_e/k_{\rm B} \sim T_e n_e $, from the ionized gas densities assuming $T_e = 7400$~K. This is a typical temperature of fully ionized gas, and in fully ionized gas, the thermal pressure will equal approximately twice the electron pressure. We find electron pressures of $7.4 \times 10^{4}$~K~cm$^{-3}$ to $3.7 \times 10^{6}$~K~cm$^{-3}$ with a median value of $P_e/k_{\rm B} \sim 2.8 \times 10^{5}$~K~cm$^{-3}$. We show these values along the upper axis of the electron density histogram in Figure~\ref{fig:fil_hist}.  In the majority of filamentary structures, comparing free-free emission with tracers at different frequencies, as shown in Figure~\ref{fig:comparecont} for example, reveals that the gas is likely fully ionized. However, with such a rich diversity of activity, sweeping statements for the Cygnus X region do not tend to hold, and there may be a rare few of the fainter filamentary structures which are no fully ionized; there the electron temperature would be lower and the thermal pressure would not be equal to $\approx 2 n_e T_e$.

Regions with $3 \times 10^{5}$~K~cm$^{-3}$ are over-pressured compared with typical values in the diffuse ISM, $P/k_{\rm B} \sim 3.8 \times 10^{3}$~K~cm$^{-3}$ \citep{Jenkins2001, Jenkins2011}. Over-pressure compared to the general ISM is expected for a region where massive stars are interacting with their environment, as massive stars create regions of high pressure that will expand. Indeed the pressures we find do coincide with a small fraction ($\sim$0.05\%) of nearly all gas surveyed by \cite{Jenkins2011} that has a large pressure ($>3 \times 10^5$~K~cm$^{-3}$) and which is more prevalent at high velocities or for regions with enhanced starlight densities.

%%%%%%%%%%%%%%%%%%%%%%%
\subsubsection{Correlation between properties of filamentary structure?}
\label{sssec:comparefilprops}

\begin{figure}
    \centering
    \includegraphics[width=0.45\textwidth]{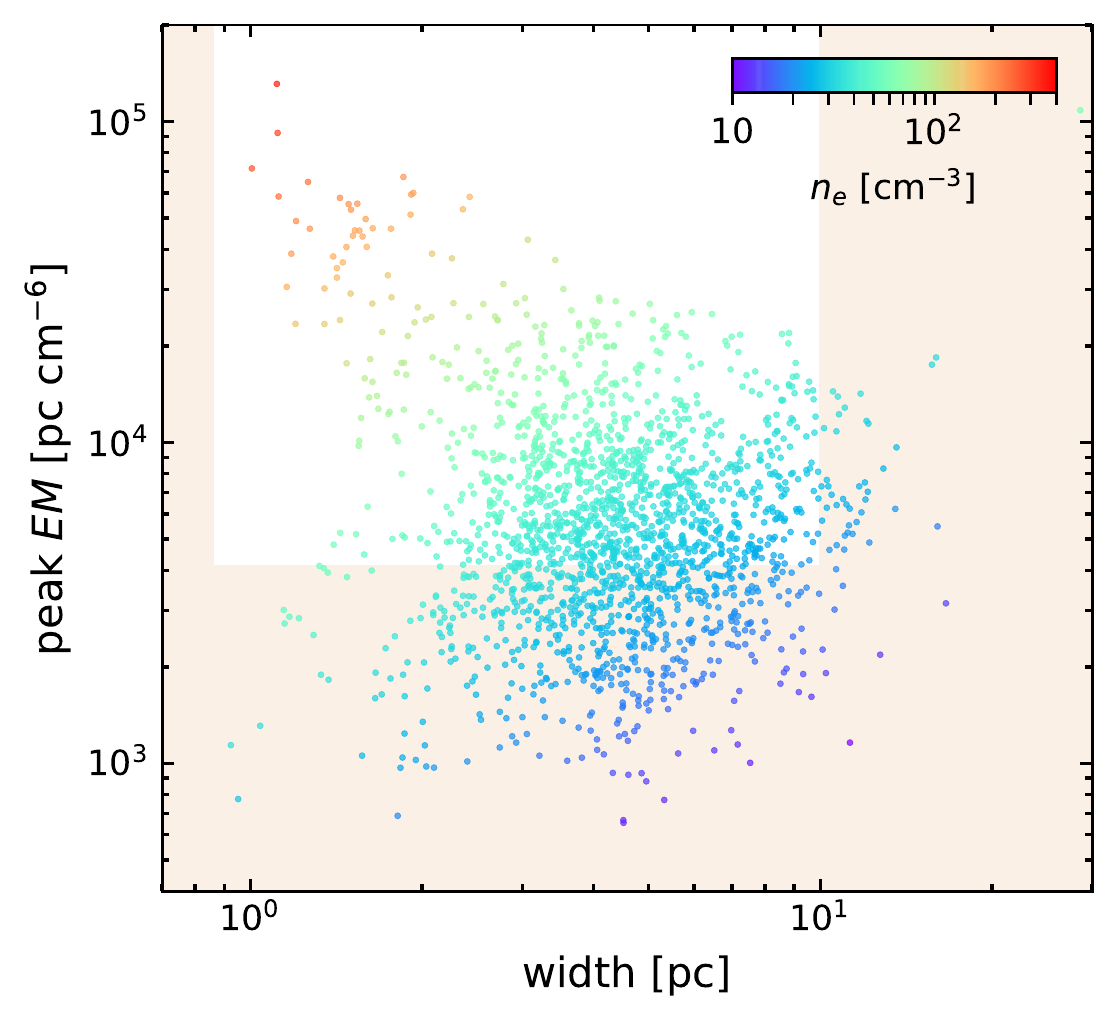}
    \caption{Peak $EM$ of the best fit to the filament radial profile plotted with the best fit FWHM. The shaded-tan areas represent values which fall outside of our completeness limits. The colors of the data points represent the electron density inferred from the filament profile $EM$ and width. 
    }
    \label{fig:fil_pw}
\end{figure}

In Figure~\ref{fig:fil_pw} we plot the peak $EM$ of the filament profiles as a function of the profile width, and we color the data points by the inferred electron density. Regions with the brightest emission measures (likely associated with classic \Hii\ regions) tend to have smaller sizes and higher densities. Whereas, emission measures of approximately $EM \lesssim 10^4$~\emuni\ (or electron densities $\lesssim 50$~\cmc) are found with a wide range of widths. While the densities are approximate -- likely accurate to within a factor of 2 -- they do follow a general trend. The distribution in the plot is curious; a dearth of high intensity {\it EM} filamentary structure with large widths would physically be harder to maintain and may represent a true de-populated portion of this plot.

%%%%%%%%%%%%%%%%%%%%%%%%%%%%%%%%%%%%%%%%
\section{Discussion}
\label{sec:discuss}

In this section we discuss the sources which maintain the ionization of the thermally emitting gas that we observe. We find Cyg~OB2 may have a considerable influence, ionizing up to two-thirds of the emission in this region. We also discuss what is forming the filamentary structure -- likely photoevaporating surfaces of neutral material, flowing into lower-density volume-filling ionized gas. While our calculations suggest that the stellar winds of Cyg~OB2 may be dissipating turbulence in the form of transitory filaments, the filamentary structure densities correlate with incident radiation and thus suggest only a minority of filamentary structures are influenced by the stellar winds of Cyg OB2. Compression of the surrounding material may also help to form filaments \citep[e.g., see][]{Zavagno2020}. We place our results in the context of ionized gas surveyed with \Nii\ FIR fine structure lines --- finding remarkable agreement with properties --- and we construct a framework in this region for ELD ionized gas and how it is maintained.  Lastly, we inform on a bright future for LOFAR observations of diffuse emission in the Galactic plane.

%%%%%%%%%%%%%%%%%%%%%%%%%%%%%%%%%%%%%%%%%%%%%%%%%%%%%%
\subsection{Source of ionization of the filamentary structure}
\label{ssec:filionsource}

%%%%%%%%%%%%%%%%%%%%%%%%%%%%%%%%%%%%%%%%%%%%%%%%%%%%%%
\subsubsection{Local massive stars}

Signatures of the youngest, most dense regions of active star formation (e.g. DR~21, W75N) are not traced by free-free emission at the low frequency of 148~MHz because of large free-free opacities. Somewhat more evolved regions, which have begun dispersing their molecular material and bursting open, have lower gas densities and smaller emission measures, as we see, leading to a more pronounced spatial separation between the star and warm gas which it ionizes. In these regions, local ($d \lesssim 10$~pc for a single O3 star) sources supply ionizing photons. Examples of this include AFGL~2636, BD+43\degree~3654 and DR~16, DR~20, DR~22, and the complex environments of and between DR~17 and DR~23. The regions named have not just one but a small cluster of massive OB stars within them. These regions are generally found at larger distances ($d \gtrsim 30$~pc) from Cyg OB2, to the east in the image. Likewise (single) massive field stars, for example to the northeast and southeast regions of the image, shape local filamentary structure. In these regions, the ionized gas tends to be bounded on one side by cold, molecular gas; the low density ionized gas does not often envelope cold gas and dust, only in select small pillar and globules.

%%%%%%%%%%%%%%%%%%%%%%%%%%%%%%%%%%%%%%%%%%%%%%%%%%%%%%
\subsubsection{Cyg~OB2}
\label{sssec:cygob2ion}

\begin{figure*}
    \centering
    \includegraphics[width=0.45\textwidth]{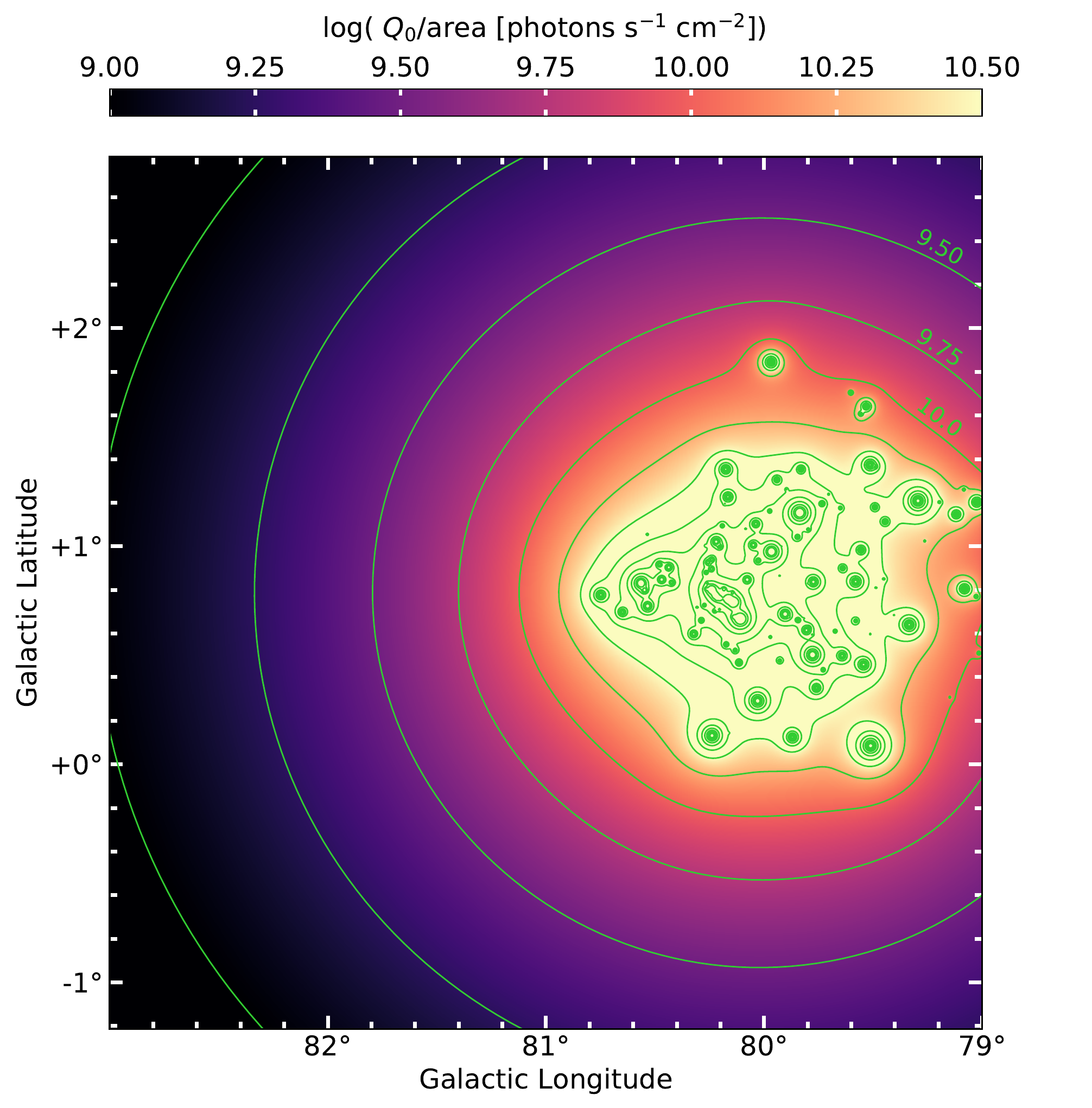}
    \includegraphics[width=0.45\textwidth]{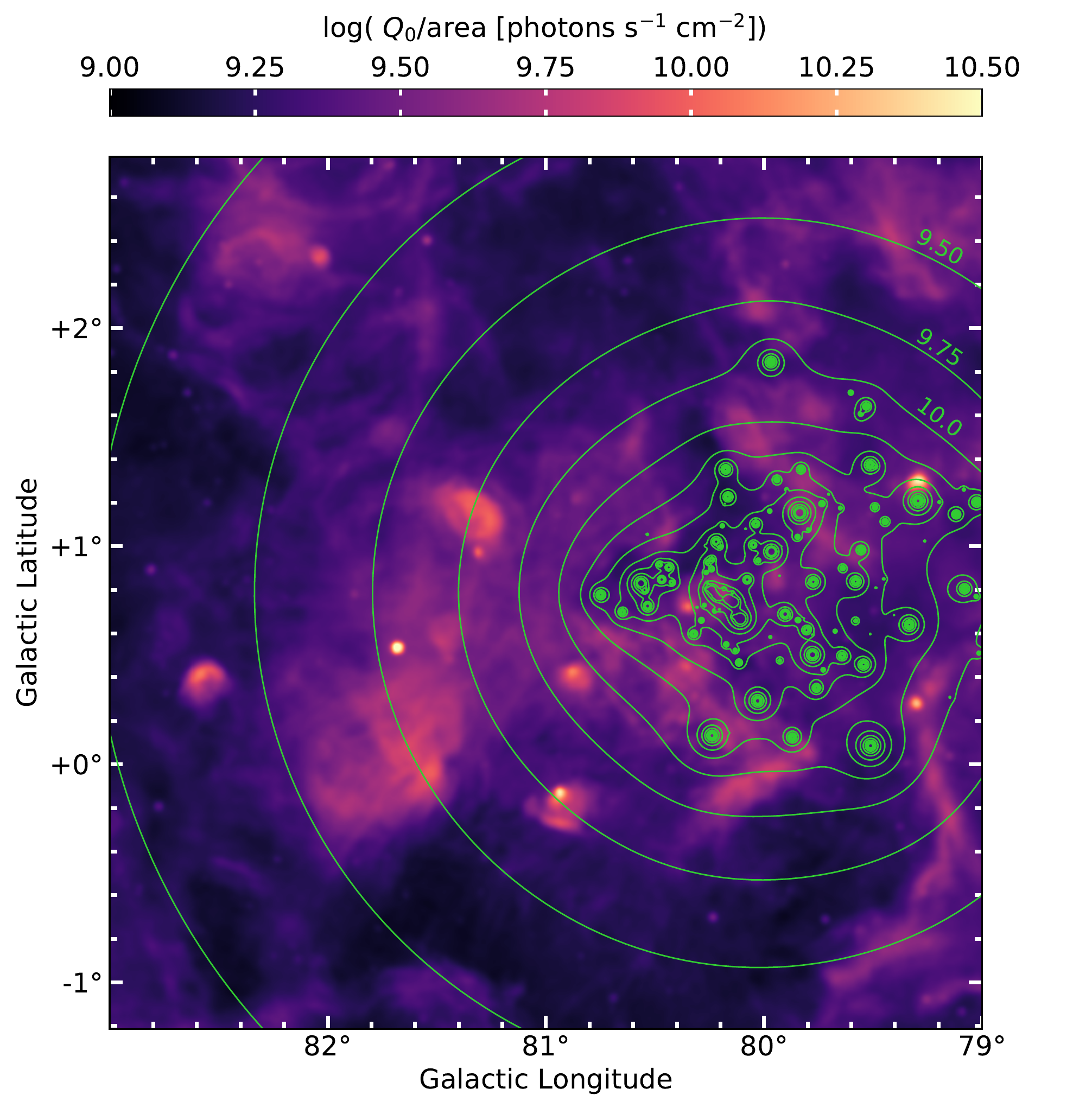}
    \caption{\textit{Left:} A map of the ionizing photon ($>13.6$~eV) rate per unit area constructed from the observed OB stars of the Cyg~OB2 association (for details see Appendix~\ref{asec:cygob2}). Green contours are shown at $\log (Q_0/{\rm area~[photons~s^{-1}~cm^{-2}]}) = (9,9.25,9.5,... 12.25)$. \textit{Right:} The ionizing photon rate per unit area as traced by thermal free-free emission. The green contours of the Cyg~OB2 ionizing field are overlaid. }
    \label{fig:cygob2_uv}
\end{figure*}

To explore the role of Cyg~OB2 in ionizing filamentary structure, we construct a map of the EUV ($>13.6$~eV) ionizing photon rate per unit area from Cyg~OB2, as shown in Figure~\ref{fig:cygob2_uv}, from catalogs of OB stars \citep[][and references therein]{Berlanas2018}. We describe how the map is created in detail in Appendix~\ref{asec:cygob2}, following the procedure of \cite{Tiwari2021}. Two important features of the map are that it is a simple 2D projection and  we do not attempt to account for any absorption or attenuation of ionizing photons. The total ionizing photon rate of the association from our model is $6 \times 10^{50}$ EUV-photons s$^{-1}$, which we note agrees with the expectation of a stellar population at zero age main sequence to 3 Myr old with stellar mass $M_{\star} \approx 1.6 \times 10^{4}$~\Msun\ \citep[Starburst99;][]{Leitherer1999}.

In Figure~\ref{fig:cygob2_uv} we also show the ionizing photon flux estimated directly from the optically thin thermal radio continuum at 1.4 GHz, from the relation \citep[e.g.,][]{Emig2020b},
\begin{equation}
\begin{split}
    Q_0/\mathrm{area} = 
    \left( 2.49 \times 10^{8}~\mathrm{photons~s^{-1}~cm^{-2}} \right)
    \left( \frac{ EM }{10^3~\mathrm{pc~cm^{-6}}} \right)  \\
    \left( \frac{ T_e }{ 7400~\mathrm{K} }\right)^{-0.833 - 0.034 \ln(T_e / 7400~\mathrm{K})}.
\end{split}
\label{eq:q0area}
\end{equation}
The Cyg OB2 ionizing photon flux outshines a large portion of the inferred flux from the radio emission, and the radio structure generally avoids direct overlap with the locations of stars in Cyg OB2. We quantify how much of the low-density ionized gas can be maintained by the ionizing photons from Cyg OB2. We take the ratio of the two images that are shown in Figure~\ref{fig:cygob2_uv}, which equates to the observed ionizing photon flux divided by Cyg OB2's output ionizing photon flux. We compute that at 67\% of the pixels in this region 
the ionizing photon flux from Cyg OB2 is sufficient to maintain the ionization.

We make a rough comparison of the influence of Cyg OB2's ionizing radiation on filamentary structure. Starting from the mask of the filament spines, we extend the mask at each filament segment to a square region of 29 pixel sides centered on the segment. This results in about a third of the pixels in the image being attributed to a filament. Taking the ratio of the images and considering only the masked pixels, we again find that for 67\% of the pixels the ionizing photon flux from Cyg OB2 is sufficient to provide the ionization.

\begin{figure*}
    \centering
    \includegraphics[width=0.33\textwidth]{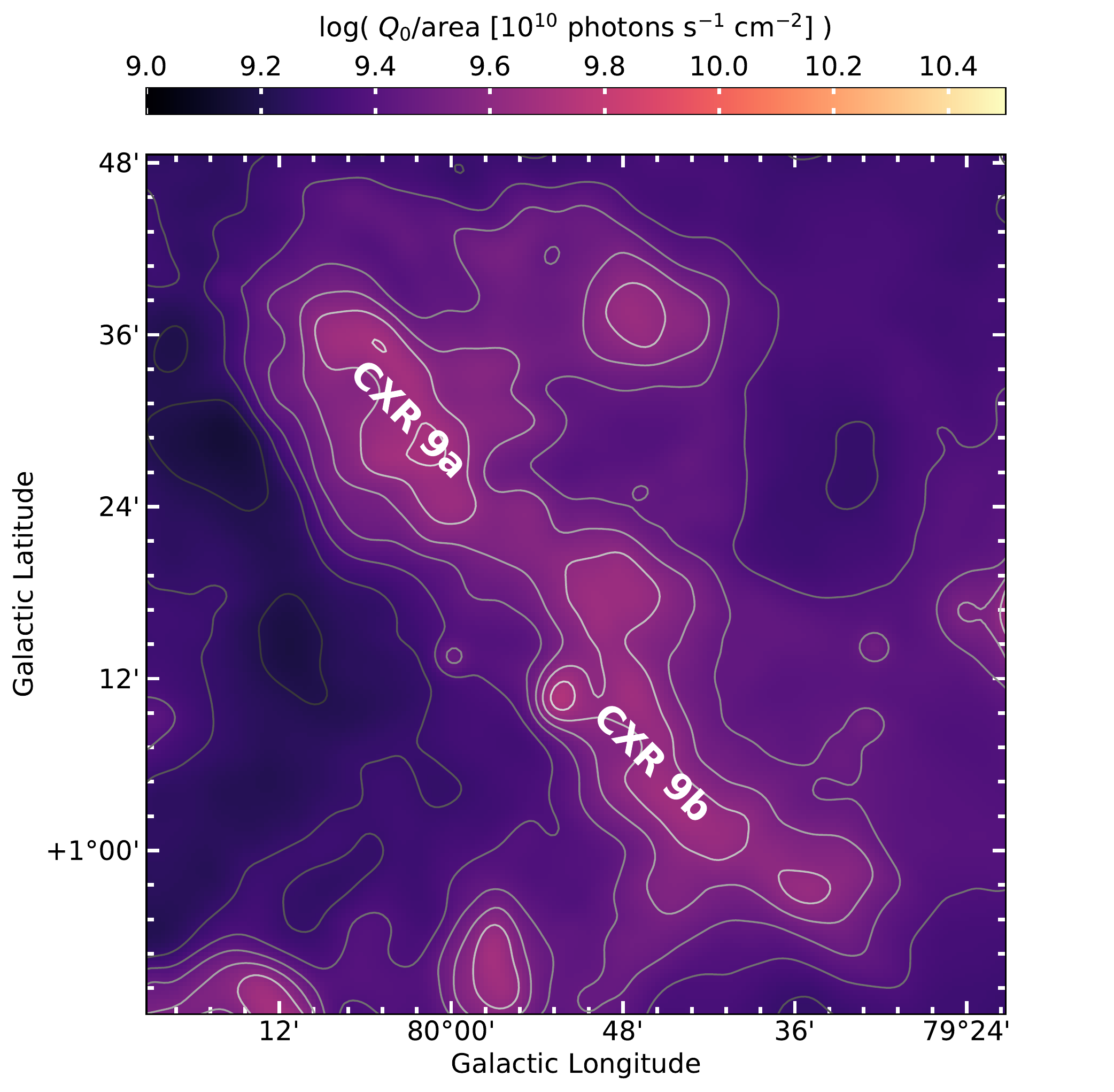}
    \includegraphics[width=0.33\textwidth]{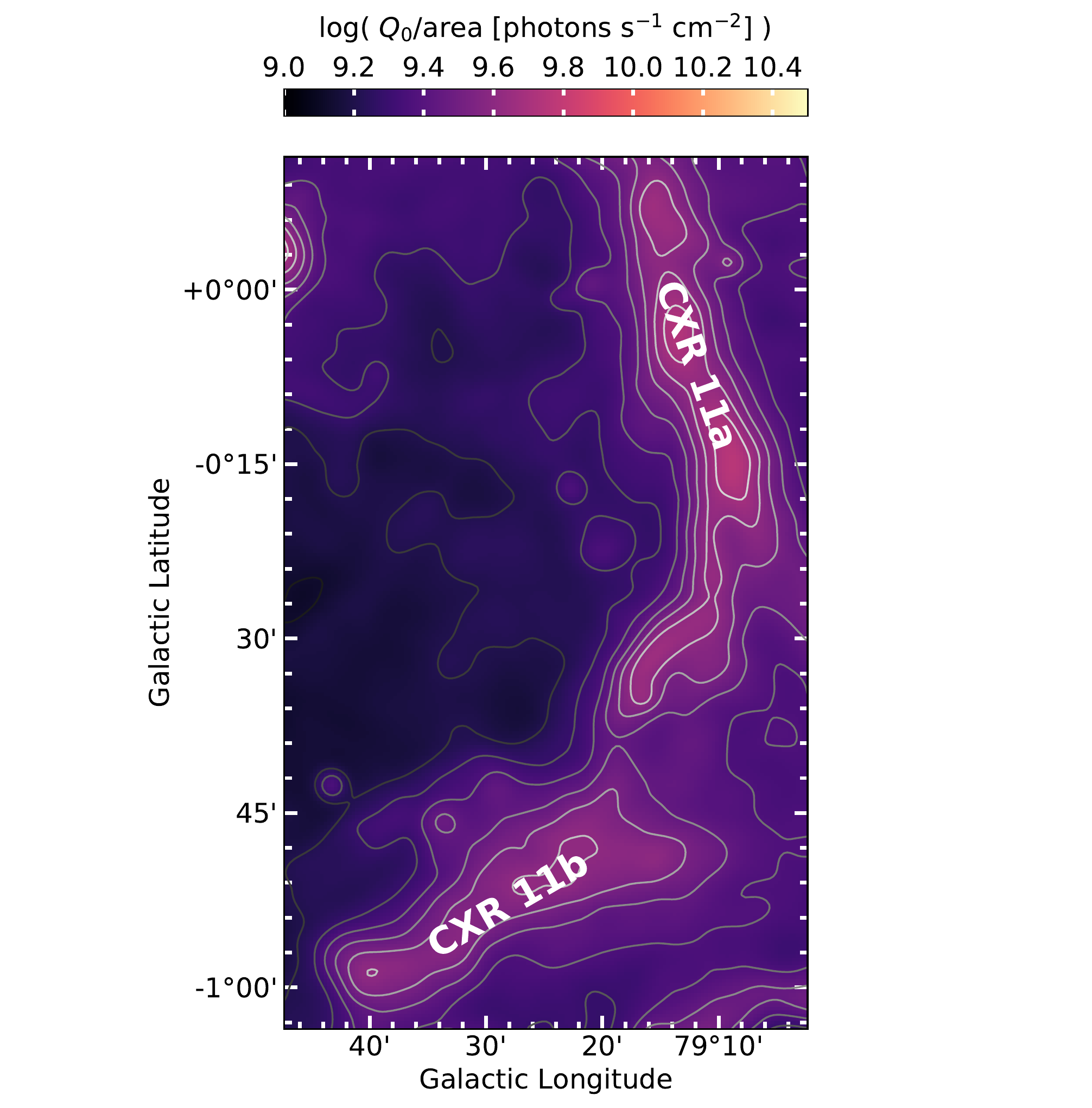}
    \includegraphics[width=0.33\textwidth]{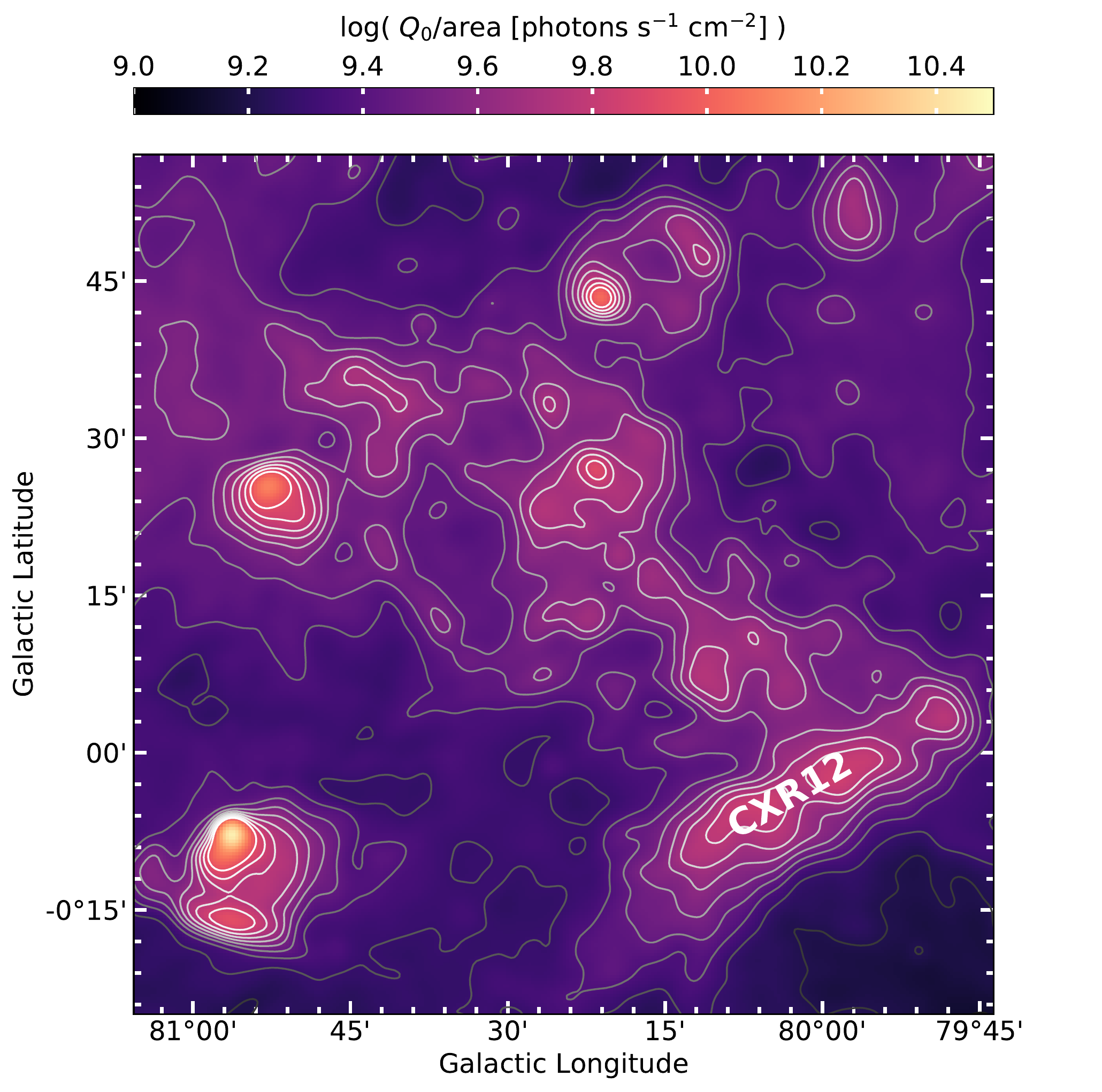}
    \caption{Here we show a zoom-in on filamentary structures whose properties we discuss in more detail (see Table~\ref{tab:filprops} and Sections~\ref{ssec:filionsource}~\&~\ref{ssec:filorigin}).}
    \label{fig:zoomfils}
\end{figure*}

We focus on some of the features and regions which appear to be influenced by Cyg OB2 in Figure~\ref{fig:zoomfils}. These radio continuum features correlate well with 24~$\mu$m emission (see Figure~\ref{fig:comparecont}) and do not appear to be directly related to individual \Hii\ regions. In Table~\ref{tab:filprops} we include estimated properties of identified filamentary structure. The widths, emission measures, and densities have been computed with the filament analysis described in Section~\ref{sec:filaments}. The length has been estimated by eye using the furthest extent of a straight line along a filament, starting and ending where the emission reaches a local intensity of half maximum. In this manner, the length of the curved features such as CXR~9a,b are slightly underestimated. We also compute the ionizing photon flux and the ionizing photon rate. In these filamentary structures, the ionizing photon flux from Cyg OB2 at their projected distances is a factor of 10 or more than required. 

We also note that apart from coming from UV radiation, the presence of ionization in these structures (especially CXR 12; see Figure~\ref{fig:zoomfils}) could come from electron collisions if shocks are present there. Bands containing shock tracers such as H and K in the near IR and 4.5 microns (Spitzer) show extended emission coincident with CXR 12. Future work exploring this as a possible physical mechanism would be worthwhile.

\begin{table*}
    \caption{Properties of select filamentary structures which are likely influenced by Cyg OB2}
    \label{tab:filprops}
    \begin{tabular}{lcccccccc}
    \hline
    Filament & Width & Length & $EM$ & $n_e$ & $Q_0$/area & $Q_0$ & $M_+$ & $\mathrm{D_{CygOB2}}$\\
     & (pc) & (pc) & (\emuni ) & (cm$^{-3}$) & ($10^9$ photons s$^{-1}$ cm$^{-2}$) & ($10^{48}$ photons s$^{-1}$) & (\Msun) & (pc) \\
    \hline
    CXR~9a & 5.3 & 8.0  & 5600 & 33  & $1.4$  & 2.8  & 740  & 19 \\ 
    CXR~9b & 4.6 & 14.9 & 5600 & 35  & $1.4$  & 2.1  & 1100 & 15 \\ 
    CXR~11a & 3.6 & 14.7 & 7500 & 46  & $1.9$  & 1.7 & 880 & 34\\ 
    CXR~11b & 4.6 & 13.8 & 4200 & 30  & $1.1$  & 1.5 & 880 & 47\\ 
    CXR~12 & 4.2 & 15.0 & 7800 & 43  & $2.0$  & 2.4  & 1100 & 22 \\ 
    \hline
    \multicolumn{9}{l}{Width is the median FWHM width of all profiles of the filament.}\\
    \multicolumn{9}{l}{Length is the by-eye estimate of the longest extent of the filament.}\\
    \multicolumn{9}{l}{$EM$ is the median peak emission measure over all profiles of the filament.}\\
    \multicolumn{9}{l}{$n_e$ is the median electron density estimated from all profiles of the filament.}\\
    \multicolumn{9}{l}{$Q_0$/area is the median of the EUV ionizing photon flux estimated from all profiles of the filament.}\\
    \multicolumn{9}{l}{$Q_0$ is the median of the EUV ionizing photon rate estimated from all profiles of the filament.} \\
    \multicolumn{9}{l}{$\mathrm{D_{CygOB2}}$ is the approximate projected distance from Cyg OB2.} \\
    \end{tabular}
\end{table*}

\subsubsection{Internal B stars?}

The ionizing photon rates of the filamentary structure exemplified in Table~\ref{tab:filprops} are $\sim2 \times 10^{48}$~EUV-photons~s$^{-1}$, a rate which could be supplied by early type B stars local to the filamentary structure. \citet{Wendker1991} argued for the ridges being internally ionized by about 3 to 10 stars of type B2 to B3. They give additional reasoning that filamentary structures ionized externally are typically smaller, strongly curved, and with shallower intensity gradients at the edges. However, searches for B stars in these (low extinction) regions have subsequently been made and can now rule out the presence of B stars with confidence \citep{Comeron2012, Comeron2020}. Furthermore, in Section~\ref{sec:filaments} we mention examples of externally illuminated ionized filamentary structure of similar length, width and curvature as those observed in the Cygnus X region, and in the next section (Section~\ref{ssec:filorigin}), we extensively discuss evidence for the (external) mechanisms responsible for the forming of filamentary structure.

%%%%%%%%%%%%%%%%%%%%%%%%%%%%%%%%%%%%%%%%%%%%%%%%%%%%%%
\subsection{Origin of the filamentary structure}
\label{ssec:filorigin}

%%%%%%%%%%%%%%%%%%%%%%%%%%%%%%%%%%%%%%%%%%%%%%%%%%%%%%
\subsubsection{Photoevaporation}
\label{sssec:photoevap}

When an ionized volume of gas is not in (pressure) equilibrium with neutral gas, an ionization front propagates into the neutral medium. With increasing distance from the ionizing source, an ionization front is preceded by a layer of swept up neutral gas and somewhat further away an H$_2$ dissociation front \citep{Elmegreen1977}. Since gas at the ionization front is over-pressured compared with the inner, low-density ionized gas, a flow occurs towards the ionizing source. The back-reaction created when the ionized gas pushes off the neutral material (the rocket effect) may eject even more mass and exert significant forces. Ionized gas with enhanced density (and intensity) is observed at this photoevaporating boundary. Photoevaporating surfaces are also referred to as ionized boundary layers and occur in champagne flows \citep{Tenorio-Tagle1979a} and blister-like \Hii\ regions. 

We sketch a rough portrait of the PDR like transition from ionized, atomic, to molecular gas at a photoevaporating surface (see \citet{Hollenbach1999} and references therein). Let's derive the width of the neutral region of the interface in order to compare the distribution and  morphology of the ionized gas (4.3~pc width) and the PAH emission. Assuming the gas in the filaments is fully ionized up to the ionization front, the median thermal pressure is $P / k_{\rm B} = 2 n_e T_e \approx 5.6 \times 10^5$~\puni. The temperature of neutral material is $T \sim 100$~K at an extinction of $A_{\rm V} \sim 2$~mag.  Assuming the ionized gas at this ionization front is in pressure equilibrium with neutral material, the gas would have a density $n_{\rm PDR} \sim 5600$~cm$^{-3}$. Adopting ${\rm N_{\rm PDR}} / A_{\rm V} = 2.1 \times 10^{21}$~cm$^{-2}$~mag$^{-1}$ \citep{Zhu2017}, then $N_{\rm PDR} / n_{\rm PDR} \approx 0.2$~pc. Thus, for these high pressures, we expect a relatively narrow region of approximately 0.2~pc (or 30\asec\ for Cygnus X) of neutral gas bright in PAH emission between the ionized gas surface and the dissociation front.
For less intense radiation or a less dense medium, the width of the neutral region extends further.

Photoevaporating boundaries are seen at the PDR interfaces traced by 8~$\mu$m emission in Figure~\ref{fig:comparecont} in more compact regions of star-formation that have local sources of intense radiation -- e.g., filamentary structure associated with sources AFGL 2636, DR~7, DR~17, DR~22 and DR~23. Other possible examples are the filamentary structures which fall in the regions between DR~16 and DR~17 -- where in Figure~\ref{fig:comparecont}, emission is found surrounding massive stars but not overlapping with them.

For gas flowing away from the ionization front approximately at the sound speed,  $C_{\rm \Hii} \sim 10$~\kms,\footnote{$c_{\mathrm{s}}=\sqrt{k_{\mathrm{B}}T/\mu m_{\mathrm{H}}}$, with $\mu$ the mean weight of the particles and $m_{\mathrm{H}}$ the mass of a hydrogen atom} a filament-like structure of 4.3~pc would be visible for 0.4~Myr, but would continue eating away at the neutral clouds. An ionized mass of 940~\Msun, which is the mean value of the Table~\ref{tab:filprops} entries, implies a mass loss rate of the neutral cloud of 2400~\Msun~Myr$^{-1}$. Each photoevaporating surface with these approximate characteristics would evaporate $1.2 \times 10^4$~\Msun\ of ionized gas over the lifetime, $\sim$5~Myr, of Cyg~OB2. This mass is comparable to the stellar mass of Cyg OB2 and about 10\% of the mass of a typical molecular cloud of $\sim$10$^5$~\Msun\ \citep{Tielens2005}.

The properties of the filamentary structure identified in Cygnus X are comparable with those of the California Nebula and the ionization front IC~434. The ionized filamentary structures of the California Nebula have $EM = 2200$~\emuni\ and FWHM size of 1.1~pc, and they arise from a single, high-velocity O star passing by and illuminating a molecular cloud some 50~pc away \citep[e.g.,][]{Bohnenstengel1976}.  The ionization front IC 434 arises from a photoevaporating flow at the surface of a molecular cloud, ionized by $\sigma$~Ori with an emission measure of $EM = 2 \times 10^4$~\emuni\ measured in H$\alpha$ and about 0.2~pc wide \citep[e.g.,][]{Ochsendorf2014}. The densities computed for the California Nebula and IC~434 assuming $n_e = \sqrt{EM / \ell}$ are $\sim$44 and 47~\cmc, respectively.

\begin{figure}
    \centering
    \includegraphics[width=0.45\textwidth]{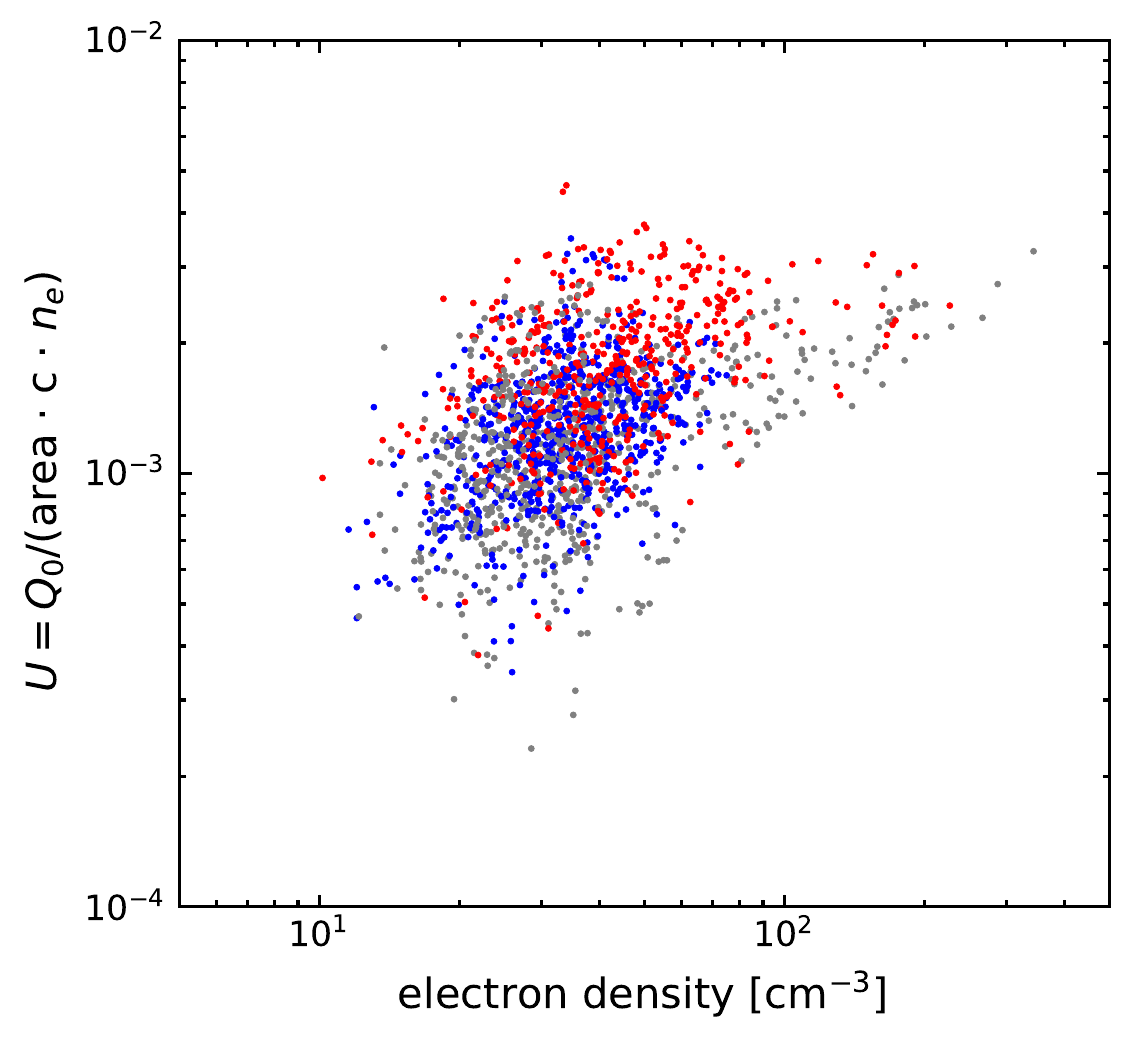}
    \caption{The dimensionless ionization parameter, $U$, estimated for each radial profile plotted against its electron density. The correlation between density and $U$ indicates that the filament properties depend on the incident radiation field. No correlation as a function of density would indicate other (feedback) processes influence the physical properties (density) of the filamentary structure. Red (blue) data points are those which fall within the red (blue) regions defined in Figure~\ref{fig:fil_loc} and correspond to the filament profiles in locations that (do not) have a local massive star that likely contributes to ionization. Typical error bars are about 0.3 dex.}
    \label{fig:Uparam}
\end{figure}

To determine if the filamentary structures show a relationship to ionizing radiation, we estimate the dimensionless ionization parameter, $U$, and plot it as a function of density in Figure~\ref{fig:Uparam}. The ionization parameter is equal to the ionizing photon flux, $\phi = Q_0 / (\rm area)$ as in Equation~\ref{eq:q0area}, divided by the density as well as speed of light,
\begin{equation}
    U = \frac{ \phi }{ n_e \, c},
\end{equation}
The ionization parameter is a measure of the ionization rate over the recombination rate. The ionization parameter is indicative of the ratio between radiation and gas pressures. The electron density is expected to depend on $U$ when its properties depend on the incident radiation field -- also including when radiation pressure has influence. The ionization parameter has typical values of $U \gtrsim (10^{-2} - 10^{-3})$ in \Hii\ regions while the WIM has $U \sim 10^{-4}$ \citep[e.g.,][]{Tielens2005, Kewley2019}. We compute $U$ by plugging-in the $EM$ of a given filament segment into Equation~\ref{eq:q0area} and dividing by the derived $n_e$ of the filament segment.

In Figure~\ref{fig:Uparam}, we also colored the data points, depending on whether a local massive star is known to the filament (and thus could be responsible for the ionization) shown in red, and regions without a clear ionization source in blue. Because there is a general trend of increasing $U$ as the density increases in our data, irrespective of whether a local star is near the filamentary structure, this indicates that ionizing radiation influences the filament properties. This would argue against stellar winds compressing and regulating the ionized gas pressure and density (see Section~\ref{sssec:winds}). 

We mention a caveat in these calculations. If the ``filaments'' are truly photoevaporating surfaces, then their emission is more likely to be distributed in sheets rather than filaments, and thus traversing a longer pathlength along the line of sight. The electron densities of the ionized gas measured in this way would be about two times smaller, if we assume that the pathlength along the line of sight is equal to the longest dimension of the filamentary structure (which seems to be at most a factor of four longer than the width). 

Looking at this scenario from a different perspective, we also discuss whether filamentary structures that do not have or are significantly displaced from PAH or $^{13}$CO emission can plausibly be photoevaporating flows. The largest filamentary structures, which are far from individual star-forming regions and may be influenced by Cyg~OB2, have significant offsets between the ionized ridge and neutral material (as traced by $^{13}$CO or 3.6,8~$\mu$m emission). Examples of this include CXR~12, CXR~11a,b, and the region just south-east of the Cyg~OB2 core --- all of which are shown in Figure~\ref{fig:zoomfils}. Using CXR~11b as an example, the projected distance between the ionized ridge and the PAH emitting ridge is $d \approx 7$\amin~$\sim$~3~pc. The large separation would imply that the neutral gas is rather diffuse; the estimated density is $n_{\rm PDR} = {\rm N_{\rm PDR}} / d \sim 450$~\cmc, since ${\rm N_{\rm PDR}} \approx 4.2 \times 10^{21}$~cm$^{-2}$. Pressure equilibrium, at $T \approx 100$~K would thus be $P / k_{\rm B} \sim 4.5 \times 10^4$~\puni. For a temperature of 7400~K, the ionized gas density is expected to be 6~\cmc. Even if we assume CXR~11b is a sheet rather than a filament and we let the pathlength along the line of sight equal the length along the plane of the sky, $\ell = 13.8$~pc, the electron density is still a factor of three too large,  $n_e = \sqrt{ EM / \ell } = 17$~\cmc. Note that this is comparable to our error estimate for the density (Section~\ref{sssec:fil_ne}). In conclusion, ionized filamentary structures that do not have neutral gas abutting the filament-like edge are likely not photoevaporating surfaces.

%%%%%%%%%%%%%%%%%%%%%%%%%%%%%%%%%%%%%%%%%%%%%%%%%%%%%%
\subsubsection{Stellar Winds}
\label{sssec:winds}

Next, we turn our attention to the strong stellar winds of Cyg~OB2. Multiple interactions of the winds from massive stars produce diffuse shock-heated X-ray gas.\footnote{Any previous (generations of) supernovae would also contribute to shock-heated X-ray gas.} A region of hot ($T \gtrsim 10^6$~K) plasma fills the volume surrounding the star(s) at small radii. Swept-up interstellar gas is compressed at larger radii \citep[e.g.,][]{Weaver1977, Harper-Clark2009, Lancaster2021a}, forming a relatively dense shell of ionized gas and warm dust that is in pressure equilibrium with the hot gas and/or which may be fractal in structure \citep{Lancaster2021b}.

A {\it Chandra} X-ray survey centered on Cyg OB2 has revealed a $R \approx 4.1$~pc cavity of hot ($T_{\rm X} \approx 5 \times 10^6$~K) and diffuse ($n_{\rm X} \approx 0.06$~cm$^{-3}$) gas \citep{AlbaceteColombo2018}. In comparison, the core of Cyg~OB2 marked with a gray circle in Figure~\ref{fig:comparecont}, has a radius of $r = 34$\arcmin~$\sim$15~pc. X-ray observations in the Cygnus X region are complicated by high levels of absorption (especially soft X-rays, which are dominated by wind-ISM interactions) in the Galactic plane; the total extent and intensity of the X-ray emission could reasonably be  underestimated. Estimating the volume averaged thermal pressure from hot gas, we find  $P_{\rm X} / k_{\rm B} \approx 2 n_{\rm X} T_{\rm X} \sim 6 \times 10^5$~\puni, where $n_{\rm X}$ and $T_{\rm X}$  are the electron number density and temperature. This matches the median thermal pressure of the ionized filamentary structures, which is what would be expected for ionized gas in pressure equilibrium with hot shocked gas. 

There are several radio filamentary structures abutting a cavity around the Cyg OB2 stars and where dense gas counterparts as traced through PAH or $^{13}$CO emission are not immediately obvious. The main region under consideration is shown in Figure~\ref{fig:zoomfils}\textit{left}; it stretches $\sim$32~pc in projection, ending in the southwest with CXR~12. As seen most clearly in Figure~\ref{fig:comparecont}, the 24~$\mu$m emission (coincident with thermal radio emission) does not appear to be bordered by PAH emission at 8~$\mu$m. $^{13}$CO avoids the regions of ionized gas (warm dust), however $^{13}$CO is present in projection in some regions at larger radii. Another region to consider is CXR~9. While faint PAH emission and traces of $^{13}$CO emission are seen towards the south, in CXR~9b, dense gas does not seem to be related to CXR~9a. The regions in consideration are close to Cyg OB2 in projection, falling within about a 20~pc radius. 

If the filamentary structures (which are not associated with neutral gas) are influenced by stellar winds, they may be short-lived turbulent features. A correlation between the luminosity of giant ($r >10$~pc) \Hii\ regions (and their size) and the line-width has been established \citep[e.g.,][]{Terlevich1981}, though the origin of the supersonic turbulence may be gravity-driven or (wind)feedback-driven. We note the ionized gas line-widths in the (resolved) filamentary structures of Cygnus X are broader than the purely thermally broadened expectation. For example, the median ionized gas FWHM line-width of the zoom-in regions of Figure~\ref{fig:zoomfils}, determined from H$110\alpha$ (4.8~GHz) observations, is $\mathrm{\Delta v} = 27$~\kms\ \citep{Piepenbrink1988}, whereas the Doppler-broadened thermal line-width\footnote{$\mathrm{\Delta v} = (30.25~\mathrm{km~s^{-1}}) \left( \frac{m_p}{m} \frac{T_e}{2 \times 10^4~\mathrm{K}} \right)^{1/2}$ where $\mathrm{\Delta v}$ is the line FWHM, $m_p$ is the proton mass and $m$ is the nuclear mass \citep{Brocklehurst1972}.} is $\mathrm{\Delta v_{th}} = 16$~\kms\ \citep{Brocklehurst1972}. The rms turbulent velocity width is calculated as $\mathrm{\Delta v_{rms}} = \sqrt{ \mathrm{\Delta v}^2 - \mathrm{\Delta v_{th}}^2 } \sim 22$~\kms.

We determine the thermal energy in the hot gas as $E_{\rm X} = P_{\rm X} (\frac{4}{3} \pi R^3) \sim 7 \times 10^{47}$~erg. Using Starburst99 \citep{Leitherer1999}, the energy output from stellar winds of an $M_{\star} = 2 \times 10^{4}$~\Msun\ population of age 4~Myr is estimated at $E_{\rm w} \sim 3 \times 10^{52}$~erg. Since the mechanical energy of the wind is more than four orders of magnitude larger than the hot gas energy and classic theory predicts it should be roughly half \citep{Weaver1977}, the hot gas may be venting into the medium and/or cooling at hot gas bubble interface may be extermely efficient \citep{Harper-Clark2009, Lancaster2021a}. Judging also by the morphology of the gas, the hot gas does not appear to be entirely well contained. 

Averaging the mechanical energy of the wind over the 4~Myr lifetime of the association, the rate of energy injection by the winds is estimated at $\dot{E}_{\rm w} \sim 2 \times 10^{38}$~erg~s$^{-1}$. Let's compare that with the calculated energy dissipation rate in the potentially turbulent-induced filament. The (kinetic) energy in the observed filamentary structure is $E_{\rm fil} = \frac{1}{2} M_+ \mathrm{(\Delta v_{rms})^2 } \sim  \frac{1}{2} (940~\mathrm{M_{\odot}}) (22~\mathrm{km~s^{-1}})^2 \sim 5 \times 10^{48}$~erg. Turbulent energy will be dissipated on a timescale comparable to the sound crossing timescale,  $t = \mathrm{width / \Delta v_{rms}} \sim (3.6~\mathrm{pc}~/~ 22~\mathrm{km~s^{-1}}) \sim 0.2$~Myr. We calculate the energy dissipation rate (potentially) provided by filaments by bringing the energy and timescales together, $\dot{E}_{\rm fil} \sim  8 \times 10^{35}$~erg~s$^{-1}$.  Hence, we conclude that only $\sim0.4$\% of the mechanical energy in the stellar winds of the Cyg OB2 association could be coupled to turbulent energy in the ionized gas.

\subsubsection{Envelopes of cold molecular filaments}

Filaments are ubiquitous in molecular clouds. Studies of the dust continuum from the \textit{Herschel} Space Observatory suggest filaments have widths of 0.01-0.4~pc \citep[e.g.,][]{Arzoumanian2011, Hacar2018}. Furthermore, filamentary structure is also prevalent in the neutral ISM \citep{Soler2020}. Could the ridges of low-density ionized gas that we observe be partially-ionized envelopes surrounding cold-dust-and-molecular filaments? We conclude no. In the majority of the brightest ionized ridges, Figure~\ref{fig:comparecont} shows no association with molecular emission. Even the fainter (likely lower density) ridges of ionized emission do not overlap with molecular emission. There is a clear spatial offset, in which the ionized gas is typically found abutting one edge of the cold molecular material. \textit{Herschel} column density maps have been created and their filaments analyzed in parts of the Cygnus region covered in this paper \citep{Hennemann2012, Schneider2016a, Schneider2016b}. \cite{Hennemann2012} examined the DR21 filaments which are among the brightest filaments in the region, and they derived widths of $\sim$0.3~pc, that may be partially affected by spatial resolution. We directly compare their FIR column density maps \citep[see also,][]{Schneider2016a} with the ionized gas emission and the spines identified by \disperse, and that shows no spatial correlation.

\subsubsection{On the observed width of the ionized emission}

Although the filament-like structure is observed to have a characteristic width that is significantly larger than our spatial resolution, \cite{Panopoulou2022} and references therein showed how this may be an artifact due to extended and confusing emission and a fractal like nature of the emission. Using FIR Herschel continuum images, \cite{Panopoulou2022} found filaments observed in the \textit{Herschel} Gould Belt Survey to have a characteristic width of about four times the resolving beam; thus more distant sources appear to have larger widths, but this is apparently resolved to smaller scales in more nearby sources.

The median width we observe is about 5 times our beam size. Although we did not focus on the fainter emission in this analysis, it does show fractal-like structure --- for example, the emission to the north east of CXR12 --- on scales smaller than the larger filament-like features we focused on. Our current analysis is limited in that we have not adequately probed smaller scale emission, and more investigation is needed to understand if the ionized emission truly has a characteristic width. Given our findings in the previous subsections of Section~\ref{sec:discuss}, it will be interesting and necessary to extend this analysis to higher resolution and deeper intensities.

%%%%%%%%%%%%%%%%%%%%%%%%%%%%%%%%%%%%%%%%%%%%%%%%%%%%%%
\subsection{Comparing properties of filamentary structure with \Nii\ findings}
\label{ssec:comparefils}

In Section~\ref{ssec:filorigin}, we noted that the filamentary-structure properties ($EM$, width, density) are consistent with those observed at photoevaporating surfaces of ionized boundary layers and stellar wind compressed ionized gas. Here we focus on a comparison of properties with ionized gas surveyed through \Nii.

Ionized gas with electron densities distributed around 35~cm$^{-3}$ appears to be consistent with ionized gas traced through the fine structure lines of \Nii\ at 122 and 205~$\mu$m with {\it Herschel} PACS \citep{Goldsmith2015}. \citet{Goldsmith2015} find a mean electron density of $n_e = 29$~cm$^{-3}$ over 96 lines-of-sight of 16\asec\ each in the Galactic midplane. While \Nii\ fine structure lines provide a reliable probe of the density of gas in the range $10 < n_e~[{\rm cm^{-3}}] < 10^3$, only densities as large as 100~\cmc\ were detected in their lines-of-sight. Detections of \Nii\ which fall just outside, $(\ell,b) = (78.1132^{\circ}, 0.0^{\circ})$, of the region we analyze resulted in an electron density of $n_e = (13.0 \pm 1.7)$~cm$^{-3}$ \citep{Goldsmith2015}. Furthermore, a lower limit of $n_e \gtrsim 30$~\cmc\ was derived from \Nii\ in the DR~21 region \citep{White2010}. Additionally, \citet{Pineda2019} derive the electron densities of 21 discrete spectral components using \Nii\ 205~$\mu$m and radio recombination lines (RRLs). They find an average electron density of 41~cm$^{-3}$ with values ranging from 8 to 170~cm$^{-3}$. The electron densities we find are in remarkable agreement and indicate that what we observe in the Cygnus X region with low-frequency radio emission may point to the types of environments which are traced by these \Nii\ surveys.

This conclusion is also supported by the \Hii\ column densities that \cite{Goldsmith2015} derives. They find \Nii\ column densities of $N({\rm N}^{+}) = (1 - 20) \times 10^{16}$~cm$^{-3}$. Using their estimated nitrogen fractional abundance $X({\rm N}+) = 2.9 \times 10^{-4}$, their column densities would translate into ionized hydrogen column densities of $N({\rm H}^{+}) = (3 - 70) \times 10^{19}$~cm$^{-3}$. As shown in Figure~\ref{fig:fil_hist}, the column densities of ionized gas found in the \Nii\ survey are consistent with the column densities of filamentary ionized gas in the Cygnus X region. 

\cite{Langer2021} used a combination of \Nii\ 122~$\mu$m and 205~$\mu$m, RRLs, and $^{12}$CO to determine electron temperature and density of ionized gas and assess its proximity to star-forming regions. They concluded that this dense warm ionized gas is located in or near star-forming regions.

%%%%%%%%%%%%%%%%%%%%%%%%%%%%%%%%%%%%%%%%%%%%%%%%%%%%%%
\subsection{Connection to ELD ionized gas}
\label{ssec:ELD}

Extended low-density (ELD) ionized gas \citep{Mezger1978} has characteristic densities of $n_e = (5 - 10)$~\cmc\ and pathlengths of (50--200)~pc. In the region we investigate, diffuse thermal emission of $EM = 5300$~\emuni\ is seen across the $\mathcal{O}$(100 pc) region.
In addition to filling the Cygnus X region, emission may also arise along the line of sight, as we are looking down a spiral arm. Thus considering path-lengths of 0.1 -- 1~kpc, the volume filling electron density is estimated at $n_e \approx (2-7)$~cm$^{-3}$. We take $n_e \approx 5$~cm$^{-3}$ as a representative value, and note this is accurate to within a factor of two. The mass of ionized gas in the $R \approx 50$~pc volume we analyze is then $6 \times 10^4$~\Msun\ (or $2 \times 10^5$~\Msun\ in a Str\"{o}mgren volume). Ionized gas of density $n_e \sim 5$~\cmc\ and cloud size $100$~pc are consistent with the properties of ELD ionized gas. Tying the volume-filling ionized gas in Cygnus X with ELD ionized gas is qualitatively consistent with previous results which connect ELD ionized gas to envelopes of \Hii\ regions \citep{Shaver1976a, Anantharamaiah1986, McKee1997} and those which suggest ELD gas predominantly arises from (less than $\sim$20 of) the most massive regions of star formation in our Galaxy --- that have cleared much of their natal molecular cloud material but are still young enough to have considerable massive stars \citep{Murray2010, Kado-Fong2020}.

The thermal pressure felt by the volume filling warm ionized gas, for an electron temperature of $T_e \approx 7400$~K and a fully ionized medium, is $P /k_{\rm B} = 2 n_e T_e \sim 7.4 \times 10^4$~\puni. While this pressure is a factor of 10 lower than in the ionized filamentary structure ($P/k_{\rm B} \sim 6 \times 10^5$~\puni) and the X-ray emitting plasma, it is still elevated compared with the diffuse ISM ($P/k_{\rm B} \sim 4 \times 10^3$~\puni) \citep{Jenkins2011}. Over-pressurization is consistent even for the globally elevated ISM pressure expected in spiral arms and at small galactic radii \citep{Wolfire2003}. 

In the Cygnus X region, ELD ionized gas may be filled and replenished by photoevaporating regions (similar in properties to \Nii\ gas of \citep{Goldsmith2015, Pineda2019, Langer2021}) eroding neutral clouds over the lifetime of the massive stars. As we calculated in Section~\ref{sssec:photoevap}, a characteristic filamentary structure in the region photoevaporates $1.2 \times 10^4$~\Msun\ of cloud mass over the lifetime of the massive stars. Five of the characteristic filamentary structures supply an equal mass of ionized gas in the region analyzed, and ten filamentary structures would replenish the full Str\"{o}mgren volume.

%%%%%%%%%%%%%%%%%%%%%%%%%%%%%%%%%%%%%%%%%%%%%%%%%%%%%%
\subsection{Future LOFAR observations}
\label{ssec:futurelofar}

Our results have demonstrated new capabilities provided by LOFAR to characterize low-density ionized gas at low radio frequencies and high spatial resolution. Future investigations in conjunction with LOFAR's LBA at 30--80~MHz will enable ionized gas (with typical ionized gas temperatures of 7000~K) to be characterized with free-free optical depths down to $\sim$400~\emuni\ \citep[e.g.,][]{DeGasperin2021}.  With the LOFAR Two Meter Sky Survey~\citep{Shimwell2019, Shimwell2022}, the Galactic Plane in the Northern hemisphere will be covered to a spatial resolution of up to 6\asec. Our analysis, builds on previous LOFAR analyses \citep[e.g.,][]{Arias2019}, which show the power of LOFAR to characterize continuum emission at low radio frequencies in the Galaxy.

%%%%%%%%%%%%%%%%%%%%%%%%%%%%%%%%%%%%%%%%
\section{Conclusions}
\label{sec:conclude}

Photoionized gas probes the influence of massive stars on their environment. The Cygnus X region ($d \sim 1.5$~kpc) is one of the most massive star forming complexes in our Galaxy, in which the Cyg OB2 association (age of 3--5 Myr and stellar mass of $M_{\star} \approx 2 \times 10^4$~\Msun) has a dominant influence. We observed the Cygnus X region at 148~MHz using LOFAR and corrected for missing short-spacing information during image deconvolution. Together with archival data from the Canadian Galactic Plane Survey, we investigate the morphology, distribution, and physical conditions of low-density ionized gas in a 4\degree~$\times$~4\degree\ ($\sim$100~pc~$\times$~100~pc) region at a resolution of 2\amin\ (0.9~pc). As first discussed at radio frequencies by \citet{Wendker1991}, the ionized gas in this region is characterized by filamentary structure. We use the \disperse\ and \filchap\ packages to characterize the radial profiles of low-density ionized filaments. Our results are as follows:

\begin{itemize}

    \item We have demonstrated a procedure for correcting for a lack of short-spacing information in LOFAR observations. We compare our results with feathering and find them to be consistent within 6\% (on average) across the region analyzed, thanks to LOFAR's excellent sensitivity to large scale emission. Future LOFAR HBA observations, especially together with the LOFAR LBA at 30--70~MHz, will characterize the low-density ISM to deep emission measures and high spatial resolution. 

    \item Radio continuum emission in the region is largely consistent with free-free thermal emission down to our LOFAR observing frequency of 148~MHz. This agrees well with previously analyses down to 408~MHz \citep{Wendker1991,Xu2013}.
    
    \item The low density ionized gas traced by the radio continuum shows a strong correspondence with warm dust traced by 24~$\mu$m emission. 
    
    \item We characterize 1874 $EM$ radial profiles from filaments. We find a power-law distribution in peak $EM$ down to our completeness limit of 4200~\emuni. A characteristic width of 4.3~pc arises in the distribution, well separated from our spatial resolution of 0.9~pc. The median electron density within the filamentary structure is 35~cm$^{-3}$ -- quite similar to the ionized gas probed along the Galactic plane through \Nii\ \citep[e.g.,][]{Goldsmith2015, Pineda2019}.  We derive thermal pressures within a median value of $6\times 10^5$~K~cm$^{-3}$, indicating the filamentary structures are over-pressured compared with the neutral ISM.
    
    \item We construct an ionizing photon map of the Cyg~OB2 association and compare it with the ionizing photon flux measured from the thermal radio continuum. We find that the ionizing photon flux from Cyg~OB2 is sufficient to maintain ionization in 67\% of the region and the filamentary structure.
    
    \item  We estimate that the (high-pressure) filamentary structures are likely photoevaporating surfaces flowing into volume-filling warm ionized gas that is relatively low in pressure (with density $n_e \sim 5$~\cmc). We notice a trend between the ionization parameter, $U$, and electron density, indicating filamentary structures are primarily influenced by stellar radiation. We estimate that typical photoevaporating surfaces influenced by Cyg~OB2 each convert $\sim 2 \times 10^4$~\Msun\ of neutral material (or 10\% of a typical $\sim 10^5$~\Msun\ molecular cloud) into ionized gas over the lifetime of the association. 
    
    \item A minority of filamentary structures do not have or are significantly displaced from neutral gas (as traced by PAH and $^{13}$CO emission) and may not be photoevaporating surfaces. We estimate the influence of stellar winds from Cyg~OB2 and find that $\sim$0.4\% of the mechanical energy in the winds may be coupled to turbulent energy in the ionized gas. In which case, some of the ionized filamentary structures may be transitory features resulting from dissipated turbulence.
    
    \item The volume-filling ionized gas in the Cygnus X region --- $M_+ \sim 2 \times 10^{5}$~\Msun\ with $n_e \sim 5$~\cmc\ and $\sim$100~pc pathlength --- is consistent with properties of extended low-density (ELD) ionized gas \citep{Mezger1978} and connects to previous findings which attribute ELD ionized gas to envelopes of \Hii\ regions and those which suggest the most massive star forming regions likely dominate ELD emission. ELD ionized gas which leaks from this largely inhomogeneous region can be replenished in mass by $\sim$10 (typical) photoevaporating surfaces over the $\sim$ 5~Myr lifetime of the OB stars. 

\end{itemize}

While this study reports on interesting new properties, it also raises many intriguing questions. Future studies of ionized gas using radio recombination lines \citep[e.g.,][]{Anderson2021} would bring useful kinematic information. Identifying and analyzing filamentary and fractal-like structure on smaller scales and extending the structural analysis on larger scales could also differentiate between and constrain turbulent properties. Further investigation into shock tracers is warranted. Finally, high-resolution radio continuum observations in Cygnus X, for example with Global View of Star Formation in the Milky Way (GLOSTAR) survey at 1\asec\ \citep[e.g.,][]{Ortiz-Leon2021}, will reveal filamentary structure in unprecedented detail.

\begin{acknowledgements}

We thank Nicola Schneider for sharing data products, and Frits Sweijen and Alex Mechev for maintaining software on the grid infrastructure and in Leiden. KLE thanks the Green Bank Observatory for hosting her as a (remote) guest during the completion of this work. The authors thank the anonymous referee for their time and efforts reviewing this manuscript.

KLE, HJAR and AGGMT acknowledge financial support from the Netherlands Organisation for Scientific Research (NWO) through TOP grant 614.001.351. GJW gratefully acknowledges the support of an Emeritus Fellowship from the Leverhulme Trust. RJvW acknowledges support from the VIDI research program with project number 639.042.729, which is financed by NWO. AGGMT acknowledges support through the Spinoza premier of the NWO.  MH acknowledges funding from the European Research Council (ERC) under the European Union's Horizon 2020 research and innovation program (grant agreement No 772663).

This paper is based (in part) on results obtained with International LOFAR Telescope (ILT) equipment under project codes \texttt{LC0\_032}. LOFAR \citep{vanHaarlem2013} is the Low Frequency Array designed and constructed by ASTRON. It has observing, data processing, and data storage facilities in several countries, that are owned by various parties (each with their own funding sources), and that are collectively operated by the ILT foundation under a joint scientific policy. The ILT resources have benefited from the following recent major funding sources: CNRS-INSU, Observatoire de Paris and Universite d'Orleans, France; BMBF, MIWF-NRW, MPG, Germany; Science Foundation Ireland (SFI), Department of Business, Enterprise and Innovation (DBEI), Ireland; NWO, the Netherlands; The Science and Technology Facilities Council, UK; Ministry of Science and Higher Education, Poland. The research presented in this paper has used data from the Canadian Galactic Plane Survey, a Canadian project with international partners, supported by the Natural Sciences and Engineering Research Council. Part of this work was carried out on the Dutch national e-infrastructure with the support of the SURF Cooperative through grant e-infra 160152. 

\end{acknowledgements}

%%%%%%%%%%%%%%%%%%%%%%%%%%%%%%%%%%%%%%%%%%%%%%%%%%



\bibliographystyle{aa}
\bibliography{Emig2022_cygx} % if your bibtex file is called example.bib

%%%%%%%%%%%%%%%%%%%%%%%%%%%%%%%%%%%%%%%%%%%%%%%%%%

%%%%%%%%%%%%%%%%% APPENDICES %%%%%%%%%%%%%%%%%%%%%

\begin{appendix}

\section{$uv$ coverage}

\begin{figure*}
    \centering
    \includegraphics[width=0.48\textwidth]{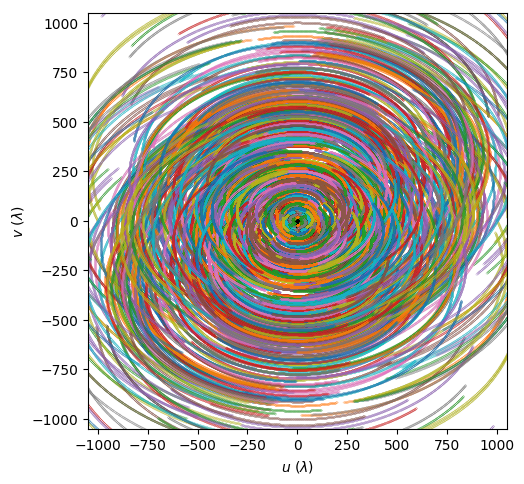}
    \includegraphics[width=0.48\textwidth]{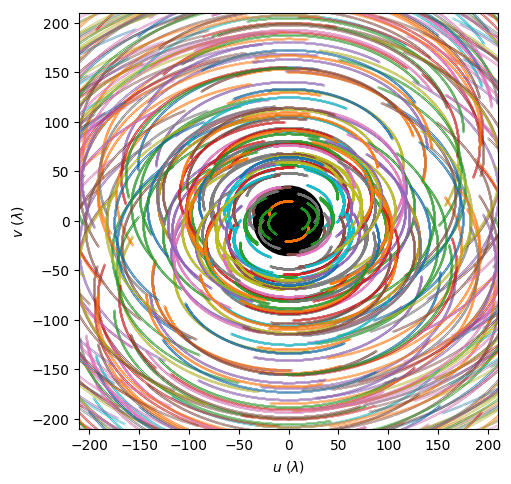}
    \caption{$uv$ coverage. \textit{Left:} The full LOFAR coverage used in this analysis. Colors depict baselines. \textit{Right:} A zoom in on the shortest baselines. The black circle encompasses the coverage provided with the 1.2\degree\ resolution short-spacing map.}
    \label{fig:uvcov}
\end{figure*}

We show the $uv$ coverage of the LOFAR observation in Figure~\ref{fig:uvcov}. At the 2\amin\ resolution of our 148~MHz image, the observations encompass maximum baseline lengths of 4200~m (equivalent to the displayed range of the abscissa in the left hand plot). The figure shows LOFAR's superb coverage to large scale emission, up to 1.6\degree. We also show the zero-spacing information that is filled in by the 1.2\degree\ resolution short-spacing map.

%%%%%%%%%%%%%%%%%%%%%%%%%%%%%%%%%%%%%%%%%%%%%%%%%%
\section{Ionizing Photons from Cyg OB2}
\label{asec:cygob2}

Using the procedure laid out in \cite{Tiwari2021}, we estimate the ionizing photon rate per unit area  from early-type Cyg OB2 cluster members by synthesizing a list of stars from catalogs by \citet{Wright2015} and \citet{Berlanas2018,Berlanas2020}, who in part used the catalog of \citet{Comeron2012}.
All stars from these catalogs have effective temperatures and luminosities assigned from spectroscopic analyses. We take the assigned parameters from \citet{Berlanas2018,Berlanas2020} if possible, then from \citet{Wright2015}, and finally \citet{Comeron2012} via \citet{Berlanas2018}.
With these effective temperatures and luminosities, we pick out models for each O or B star from the  Potsdam Wolf-Rayet (PoWR) stellar atmosphere grids \citep{Hamann2004, Todt2015, Sander2015, Hainich2019}. These models are gridded by effective temperature and gravity, so we first interpolate the gravity, log~$g$, from the grid's associated temperature and luminosity values.
From the synthetic spectra provided by the PoWR models, we integrate the total ionizing flux above 13.6~eV. We then use the stellar coordinates and the ionizing photon flux to estimate the ionizing photon rate per unit area at any location around the \Hii\ region using projected distances at 1.5~kpc and summing over the flux from all stars.

\end{appendix}

%%%%%%%%%%%%%%%%%%%%%%%%%%%%%%%%%%%%%%%%%%%%%%%%%%

\end{document}